\documentstyle[epsf,epsfig,preprint,aps]{revtex}

\tighten

\begin{document}

\bibliographystyle{unsrt}

\draft
\date{\today}
\title{
\normalsize
\mbox{ }\hspace{\fill}
\begin{minipage}{5cm}
UPR-866-T\\
FERMILAB-Pub-99/368-T
{\tt hep-th/yymmxxx}
\end{minipage}\\[5ex]
{\Large \bf 
Vacuum Domain Walls in  D-dimensions:
Local and Global Space-Time Structure}}


\renewcommand{\theequation}{\arabic{section}.\arabic{equation}}

\author{ Mirjam Cveti\v c$^a$ \footnote{ E-mail:cvetic@cvetic.hep.upenn.edu}
 and Jing Wang$^b$\footnote{E-mail:jingw@fnal.gov} }



\address{$^a$ Department of Physics and Astronomy \\ 
University of Pennsylvania, Philadelphia PA 19104-6396, USA\\
$^b$Theory Division, Fermi National Accelerator Laboratory\\
P.O.Box 500, Batavia, IL 60510, USA\\
}

\maketitle

\begin{abstract}
We  study local and global  gravitational effects
 of (D-2)-brane configurations (domain walls) in the vacuum of D-dimensional space-time.
We focus on  infinitely thin  vacuum domain walls with
arbitrary cosmological constants on either side of the wall.
In the comoving frame of the wall we derive a  general
 metric {\em Ansatz\/}, consistent with the homogeneity and isotropy of  
the space-time intrinsic to the wall, and employ Israel's matching conditions at the wall.
The space-time, intrinsic to the wall, is that of (D-1)-dimensional
Freedman-Lema{\^ i}tre-Robertson-Walker universe (with $k=-1,0,1$) which 
has a (local) description as  either  anti-deSitter, Minkowski  
or de~Sitter space-time.   For each of 
 these geometries,  we provide a systematic classification of the 
local and global space-time  structure transverse to the walls, for those with
both positive and negative tension; they   fall into
different classes according to the values of their energy density relative to
that of the  extreme (supersymmetric) configurations.
We find that in  any dimension D,  both local and global space-time structure
for each class  of domain walls  is {\it universal}.  
We also comment on the phenomenological implications of these walls in the 
special case of D=5.  
\end{abstract}

\addtocounter{footnote}{-\value{footnote}}

\newpage


\section{Introduction}

Recent months have witnessed a resurgence in the study of domain walls with 
asymptotically anti-deSitter space-times (AdS). This renewed interest 
is motivated  both  from the point of view of AdS/CFT  correspondence, providing
new insights in the study of RGE flows (see e.g., 
\cite{010,190,180,220,020,150,160,030,230,BC,deBVV}  and references therein) as
well as  from the phenomenological perspective, providing a possible resolution to
hierarchy problem in the context of  a world on a domain wall in
D=5 asymptotically AdS space-times (see, e.g.,  
\cite{050,070,060,240,250,260,211,261,081} and references therein).
This is an exiting period when formal theoretical developments
drive  phenomenological implications and vice versa.

A prerequisite for  addressing physics implications of such configurations
is a detailed understanding of their space-time structure.
 Earlier work 
 on the subject concentrated
on the domain walls in D=4 space-time dimensions,
where either side of the wall 
was to be interpreted 
as that of our four-dimensional Universe.
The first example of static domain walls between vacua with different
cosmological constants were found is \cite{CGRI} as supersymmetric
(BPS-saturated/extreme) walls interpolating between supersymmetric vacua of
D=4 $N=1$ supergravity vacua.
In  \cite{CGI} a classification of the possible supersymmetric walls
has been given, and the global
structure of the space-times induced by these walls has been
explored in \cite{CDGS,Gibb}.
A subsequent systematic  investigation \cite{CGSII} 
of the space-times of domain walls
separating regions of non-positive
cosmological constant in Einstein's theory of gravitation, 
revealed a more general
class of domain wall configurations.\footnote{Space-time properties
of non-static domain walls in D=5 were recently addressed in
\cite{170,171,Binetruy,Tye,Csaki,Kraus,Sasaki,Garriga} and references therein. 
As the results
of this paper demonstrate, the space-time structure of
 domain walls in $D$-dimensions (including $D=4$) is 
completely parallel.}
(For a review, see  \cite{CS}.)

The purpose of the present work is to generalize the  results 
of the  study of vacuum domain walls in D=4 \cite{CGSII} to
(D-2)-brane-configurations in D-dimensions, with 
 D=5  being of
special interest to the physics-implications for  the four-dimensional  
domain wall-world (as well as of  theoretical interest for   RGE flows of strongly
coupled gauge-theories in four dimensions).

Following the work of \cite{CGSII} we  derive  the local and global
properties of the space-times induced by vacuum domain 
walls in D-dimensions  between vacua of
arbitrary cosmological constant.
We start with the {\em Ansatz\/}
that the gravitational field inherits the boost symmetry of the source,
but we assume nothing about the topology
of the (D-1)-dimensional space-times parallel
to the surface of the domain wall. The space-time intrinsic to the wall, 
 are Freedman-Lema{\^ i}tre-Robertson-Walker (FLRW) universes describing locally
 (D-1)-dimensional space-times with 
 anti-deSitter (AdS$_{D-1}$), Minkowski (M$_{D-1}$) or de~Sitter
 (dS$_{D-1}$) space-times.  For each of these space-times internal to the
 wall,  the  space-time transverse to the wall can  be classified according
 to the values of cosmological constants  $\Lambda_{1,2}$ on either side of the wall and
 their relationship to the energy density of the wall $\sigma$.

 An  important result of the analysis is that the 
 space-times of domain walls  have the same universal structure in
  all D; the space-time intrinsic to the wall  is that of (D-1)-dimensional 
  FLRW universe, and the  metric coefficient, specifying the space-time
  transverse to the wall has the same form  with 
parameters  depending  on  $\Lambda_i/[(D-1)(D-3)]$.

The paper is organized as follows.  In Section \ref{grsection}
 we derive
 the line-element Ansatz,  by working
 in the comoving frame of the wall in D-dimensions. 
 In Section \ref{classification} we classify the 
domain wall solutions according to their energy-density and the values of 
 cosmological constants on either side.
We present the discussion of the result in Section\ \ref{discussion} 
, including the implications for their non-extreme
generalizations.  In particular, we discuss the examples in  $D=5$, which 
have been recently studied intensively.

\section{local properties of domain wall space-times}
\label{grsection}

\addtocounter{equation}{-\value{equation}}

In this section we present the metric Ansatz for 
vacuum domain walls
for Einstein gravity in D-dimensions;
these walls,  created from a
scalar field source,  separate
vacuum space-times of zero, positive,
and negative cosmological constants.
We study in detail only infinitely thin domain walls,
by employing Israel's
formalism\cite{ISR} of singular hypersurfaces.

\subsection{Metric Ansatz}

We solve D-dimensional  Einstein's gravitational field equations 
in the co-moving
frame of the domain wall, by assuming the following symmetry of the metric
Ansatz \cite{CGSII}:
\begin{itemize}
\item
The spatial part of the metric intrinsic to the wall is
{\em homogeneous\/}  and {\em isotropic.\/}
\item
The space-time
section transverse  to the wall is {\em static.\/}
\item
The directions parallel to the wall
are {\em boost invariant\/} in the strong
sense.
\end{itemize}

Homogeneity and isotropy reduce the 
 metric part, intrinsic to the wall,  to be 
  the spatial part of a (D-1)-dimensional
Friedmann-Lema{\^{\i}}tre-Robertson-Walker (FLRW)
metric~\cite{MTW} of the form
\begin{equation}
(ds_{\parallel})^2=R^2\left[(1-kr^2)^{-1}dr^2+r^2 d\Omega_{D-3}^2\right] ,
\label{dstwo}
\end{equation}
where $R$ is independent of the  radial coordinate 
$r$ and  the angular coordinates $\phi_i$ ($i=1,\cdots , D-3$)
 specifying the line element
$d\Omega_{D-3}^2$ of  (D-3)-sphere $S^{D-3}$.
The scalar curvature of this surface is equal to $2k/R^2$.

The sign of $k$ determines the wall geometry. 
$k=0$ defines a planar wall, in which case
the metric (\ref{dstwo}) can be written in Cartesian coordinates
$(ds_{\parallel})^2 = R^{2} (\sum_{i=1}^{D-2}dx_i^2)$.
$k>0$ corresponds to a  spherical wall-- closed bubble-- with 
$r$ and $\phi_i$'s  
 being compact coordinates; 
through a coordinate transformation $r=k^{-1/2}\sin\!\theta$
and rescaling of $R$, the line-element can be written as  
$(ds_{\parallel})^2=R^2 ( d\theta^2+\sin^2{\!\theta}
d\Omega_{D-3}^2)=R^2d\Omega_{D-2}^2$.
$k<0$ corresponds to the negative curvature, non-compact
Gauss-B{\'{o}}lyai-Lobachevski
surface; introducing  $r = (-k)^{-1/2}\sinh\!\varrho$,
with $\varrho>0 $, and rescaling $R$
yields 
$(ds_{\parallel})^{2} = R^2 (d\varrho^{2} +  \sinh^2\!\varrho
\,d\Omega_{D-3}^{2})$.

The condition that the two-dimensional space-time $(t,z)$
transverse to the wall is  {\em static\/} (as observed in the 
{\em rest frame\/} of the wall) implies the following form of 
the transverse part of the metric:
\begin{equation}
(ds_{\perp})^2=A(z)\left(dt^2-dz^2\right) ,
\end{equation}
where $z$ is the spatial direction transverse to the wall, $t$ is the 
{\em proper} time and $A(z) > 0$.
With $ds^2\equiv (ds_{\perp})^2-(ds_{\parallel})^2$ the metric takes the form:
\begin{equation}
ds^2=A(z)\left(dt^2-dz^2\right)-R(t,z)^2\left[(1-kr^2)^{-1}dr^2+r^2d\Omega_{D-3}^2
\right] .
\label{Ansatz}
\end{equation}
in which $z\in\langle -\infty,
\infty\rangle$, and the other coordinates are those of the 
FLRW cosmological model \cite{MTW}.


A straight-forward calculation of the nonzero components of the
Einstein tensor $G^{\mu}_{\;\;\nu} = {\cal R}^{\mu}_{\:\:\nu}
- {1\over 2} {\cal R}^{\alpha}_{\;\;\alpha} g^{\mu}_{\;\; \nu}$ yields the
result:  
\begin{equation}
  \left.\begin{array}{ccl}
G^{t}_{\;t}&=& \frac{D-2}{A} 
\left[ - \frac{R''}{R} + \frac{H R'}{2 R} \right] 
+ \frac{(D-3)(D-2)}{2R^2}
\left[ k + \frac{\dot{R}^2 - R^{'2}}{A} \right] \\
{\mbox{ }}& & {\mbox{ }}\\
G^{z}_{\;t}&=& \frac{D-2}{A} \left[ 
{{\dot{R}'}\over{R}}-{{H\dot{R}}\over{2 R}} \right]\\
{\mbox{ }}& & {\mbox{ }}\\
G^{z}_{\;z}&=&  \frac{D-2}{A} 
\left[ \frac{\ddot{R}}{R} - \frac{H R'}{2 R} \right]
+ \frac{(D-3)(D-2)}{2R^2} 
\left[ k + \frac{\dot{R}^2 - R^{'2}}{A} \right] \\
{\mbox{ }}& & {\mbox{ }}\\
G^{r}_{\;r}&=&G^{\phi_i}_{\;\;\phi_i}= \frac{3-D}{A} 
\left[ {{R''}\over{R}} - {{\ddot{R}}\over{R}} \right] 
 -{{H'}\over{2A}} - \frac{(D-3)(D-4)}{2R^2} 
\left[ k + \frac{\dot{R}^2 - R^{'2}}{A} \right] 
         \end{array}
  \right.\label{einst}
\end{equation}
where $i=1\cdots D-3$, $\dot{R} \equiv \partial_{t}R(t,z)$, 
$R'\equiv \partial_zR(t,z)$ and $H \equiv \partial_{z}\ln A(z)$.

The symmetry of the  matter source (as specified in the rest 
frame of the wall)
implies that the energy-momentum tensor is static, 
 with   $T^{ z}_{\;\; t}=0$ (no energy
flow in the $(t,z)$-plane)  and $T^{\mu}_{\;\;\mu}=g_\mu(z)$
 ($\mu=(t,z,r,\phi_i)$), where  $g_\mu(z)$ are functions of
$z$, only.  Then  the Einstein's equations  
$G^{\mu}_{\;\; \nu}=\kappa_D T^{ \mu}_{\; \; \nu}$ imply the following
constraints on $R(t,z)$ and $A(z)$ \footnote{ 
$\kappa_D$ is defined as $\kappa_D\equiv 8\pi G_D$, where $G_D$ is Newton's
constant in D-dimensions. We define the Lagrangian density as
${1\over \kappa_D}(-{R\over 2}+ {\cal L}_{matter}).$}: 
\begin{equation}
\left.\begin{array}{ccl}{{\dot{R}'}\over{R}}-{{H\dot{R}}\over{2 R}}&=&0, 
\label{ti1}\\
{\mbox{ }}& & {\mbox{ }}\\
\frac{R''}{R} - \frac{H R'}{2 R}&=&f_2(z),\label{ti2}\\
 {\mbox{ }}& & {\mbox{ }}\\
 {{R''}\over{R}} - {{\ddot{R}}\over{R}}&=&f_3(z),\label{ti3}\\ 
 {\mbox{ }}& & {\mbox{ }}\\
\frac{k}{R^2} + \frac{\dot{R}^2 - R^{'2}}{AR^2}&=&f_4(z),
\label{ti4} 
\end{array}
  \right.
  \label{ti}
\end{equation}
where $f_{1,2,3}(z)$ are arbitrary functions of $z$.

The static metric Ansatz ${\dot{R}}=0$ automatically satisfies Eqs. (\ref{ti}).
 For non-static metric, time-
integration of the first Eq. in 
(\ref{ti}) yields the condition 
\begin{equation}
R'=\frac{HR}{2} + f_1(z)
\label{ztint}
\end{equation}
with  $f_1(z)$  an arbitrary function of $z$. Adding $z$-derivative of
(\ref{ztint}) to  the second Eq.  in  (\ref{ti}) yields a condition $f_{1}(z)'= [-{H' \over 2}+f_2(z)]R$, which holds for any $t$ only if $f_1(z)=f_0$ is a constant and  $f_2(z)={H' \over 2}$. 

The assumption of  boost-invariance along the surfaces of constant $z$ 
implies that $f_0=0$ and as a consequence, Eq.(\ref{ztint}) is solved by 
$R^2(r,t)=A(z)S^2(t)$.
(In the static case, i.e. $\dot{R}=0$, the same symmetry
constraint also  implies $R^2(z)\propto A(z)$.)\footnote{
Namely, we assume that the gravitational field
inherits the symmetry of the source; the directions parallel to the wall are
boost invariant in the strong sense and thick walls will have
 the same $S(t)$ as found in
the thin wall approximation and will asymptotically approach the 
thin wall result for $A(z)$. (See \cite{CGSII} for a more detailed
discussion.)}

Therefore, the metric Ansatz  takes the form:
\begin{equation}
ds^2= A(z)\left\{ dt^2-dz^2-S^2(t)
\left[(1-kr^2)^{-1}dr^2+r^2 d\Omega_{D-3}^2\right]\right\},
\label{ansatz}
\end{equation}
which  has a universal form for any
D-dimensions, and thus the same  structure
as the one  obtained in D=4 \cite{CGSII}. This
 result is due to the fact that  in the wall comoving frame 
 the  homogeneity and isotropy of space-time internal to the
 wall severely restrict the form
 of the   Einstein tensor to be that of Eq. (\ref{einst}); 
 the static space-time transverse to the wall and 
 the boost invariance along the wall further fixes the metric to  be of 
 the  universal, D-independent form (\ref{ansatz}).

In the following Subsection we solve
 the Einstein equations for $A(z)$ and $S(t)$, which  specify the possible
    space-time structure  away from the domain wall.
We then employ the infinitely thin wall approximation \cite{ISR} 
in order to determine the  energy
density of the wall in terms of the  parameters in the metric.

\subsection{Local space-time solutions}  \label{subsubvacuum}

We now solve Einstein's equations for $S(t)$ and $A(z)$ 
of 
the metric Ansatz (\ref{ansatz}).
We consider thin domain walls interpolating between
two maximally symmetric vacua of zero, positive, or
negative cosmological constant.%
\footnote{ 
Maximally symmetric vacuum solutions
are well known \cite{MTW}; nevertheless,
we summarize the results here for  the 
 comoving coordinate system of the wall. Note also that 
 Israel's matching conditions 
are easily satisfied in this frame.}

Plugging the metric Ansatz (\ref{ansatz}), i.e. 
$R^2(t,z)=A(z)S^2(t)$ in the Einstein tensor (\ref{einst}),
 yields the  following equations for $S(t)$ and $A(z)$:
\begin{equation}
{\ddot{S}\over  S}=q_{0}={\dot{S}^2\over S^2}+{k\over S^2} ,
\label{seq}
\end{equation}
\begin{equation}
{1\over 4} \left({A'\over
A}\right)^2=q_{0}-\frac{\Lambda}{(D-3)(D-1)} A ,
\label{aeq}
\end{equation}
where  we have assumed that away
 from the wall, the energy momentum tensor 
 is given by $T_{\;\;\nu}^{\mu}=\Lambda \delta^\mu_{\;\;\nu}$ 
 with $\Lambda$ the cosmological constant on either side of the wall. $q_0$
  is an integration constant  satisfying
   the  consistency constraint $\Lambda A(z) \le (D-3)(D-1) q_{0} $.
 
Since the equation for $S(t)$ is independent of dimensionality, 
 the space-time  intrinsic to the wall is universal.
 Eq. (\ref{aeq}) for  $A(z)$, i.e. the metric
 coefficient  specifying the space-time transverse to the wall,  
 is  also of the same
  form   as that obtained in D=4,  except for the  D-dependent 
 coefficient in front of the 
  cosmological constant. We choose  to parameterize the
 cosmological constant as: 
\begin{equation}
\Lambda \equiv \pm (D-3)(D-1)\alpha^{2} \equiv \pm \Delta \alpha^2.
\label{lambda}\end{equation}
Thus, in terms of the parameter  $\alpha$, Eq. (\ref{aeq}) has a
universal, D-independent form and thus  yields 
the same solutions  as the ones
obtained in D=4 \cite{CGSII}.

The fact that the  domain wall space-time structure is universal, was 
anticipated in \cite{CS}. Nevertheless, the  result  is  intriguing;
{\it both the local as
well as the global space-time properties of domain walls  ((D-2)-configurations)
in  diverse (D) dimensions are universal.}
  The study of  the  local and global 
domain wall space-times in D=4
\cite{CGSII} can therefore be extended in a straightforward way to the study of
(D-2)-configurations in D-dimensions. (The special case of D=5 recently 
attracted much attention due to its phenomenological implications.)

For the sake of completeness, we shall now write down 
the  explicit results for $S(t)$
and $A(z)$ \cite{CGSII}.  
We parameterize the curvature constant of the space-time internal to the wall 
(see discussion at
the beginning of the previous subsection) as $k\in\{-\beta^2,0,\beta^2\}$.
 The solutions of $S(t)$ from Eq.\ (\ref{seq}) can be 
 classified according to the sign of $q_{0}$: 
\begin{eqnarray}
q_0=-\beta^2: \ \ S_{-} &=& \cos (\beta t 
)\;\;\;\;\;\;\; k=-\beta^2,
\label{smin}\\
q_0=0:\ \ S_{0}\; &=&  \left\{ \begin{array}{ll}
          \beta t 
	  \;\;\;\;\;\;\;\;\;  \;       &
	   k=-\beta^2, \\
             1
	                  & k=0,
           \end{array}
    \right.
\label{snull}\\
q_0=\beta^2:\ \ S_{+} &=&  \left\{ \begin{array}{ll}
\sinh (\beta t 
)\;\;\,\,  &  k=-\beta^2, \\
 e^{\beta t
 }     &  k=0 ,       \\
 \cosh (\beta t
 ) &  k=\beta^2,
\end{array}
\right.
\label{splus}
\end{eqnarray}
where the subscripts ($-,0,+$) refer to the respective  sign of $q_0$.
(Due to the  time-translation
 and time-reversal invariance,  the integration constants 
 are adjusted to yield the canonical
form for $S(t)$, and without loss of generality, 
$\beta \ge 0$ is chosen.)
The form of the solutions $S_{-,0,+}$ (Eqs.(\ref{smin})-(\ref{splus})) 
 implies that the space-time
intrinsic  to the wall is that of    AdS$_{D-1}$, M$_{D-1}$
  and dS$_{D-1}$, respectively.

Solutions of (\ref{aeq}), classified according to the sign of $q_0$,
 yield the following form of $A(z)$:
\begin{eqnarray}
q_0=-\beta^2: \ \ A_{-} &= & \begin{array}{ll}
 \beta^2 [ \alpha \cos (\beta z+\vartheta ) ]^{-2}
&
\;\;\;\;\;\;\;\;\;
\;\;\;\;\;\;
\Lambda =-\Delta \alpha^2 
\leq -\Delta \beta^2,
           \end{array}
\label{amin}\;\;\;  \\
q_0=0:\ \ A_{0}\;&=&\left\{ \begin{array}{ll}
  ( \alpha z\pm 1)^{-2}
&
\;\;\;\;\;\;\;\;\;\;
\;\;\;\;\;\;\;\;\;\;
\;\;\;\;\;
\;
\Lambda = -\Delta \alpha^2,\\
1     &
\;\;\;\;\;\;\;\;\;\;
\;\;\;\;\;\;\;\;\;\;
\;\;\;\;\;
\;
\Lambda =0,
          \end{array}
\right.
\label{anull}\\
q_0=\beta^2:\ \ A_{+}&=&\left\{ \begin{array}{ll}
\beta^2 [\alpha \sinh(\beta z-\beta z')]^{-2}
 & \;\;\;\;\;\;\,\, \Lambda = -\Delta\alpha^2,\\
e^{\pm 2\beta z}
 & \;\;\;\;\;\;\,\, \Lambda =0,\\
\beta^2 [\alpha \cosh(\beta z-\beta z'')]^{-2}
 & \;\;\;\;\;\;\,\,  \Lambda =\Delta \alpha^2\leq \Delta \beta^2,
\end{array}
\right.
\label{aplus}
\end{eqnarray}
Again, the subscripts ($-,0,+$) for  $A(z)$ refer to the sign of $q_{0}$.
Without loss of generality we have moved the origin of the $z$-axis
to the position of the wall ($z_{0}=0$).
The three integration constants
$\vartheta$, $z'$, and $z''$ are
determined by the requirement that we choose $A(z_0=0)=1$ which yields:
\begin{eqnarray}
\vartheta_{\pm}&=&\pm {\mbox{arccos}} (\beta/\alpha),\\
\beta z'_{\pm} &=&
\frac{1}{2}\ln \left[1+ {2\beta^2\over\alpha^2}\pm {2\beta\over\alpha^2}
(\alpha^2+\beta^2)^{1/2}\right],\label{deltadet}\\
\beta z''_{\pm} &=&
\frac{1}{2}\ln\left[-1+ {2\beta^2\over\alpha^2}\mp {2\beta\over\alpha^2}
(\beta^2-\alpha^2)^{1/2}\right]. \label{gammadet}
\end{eqnarray}
The constants $\beta z'$ and $\beta z''$ satisfy
$e^{2\beta (z'_{+}+z'_{-})}=
e^{2\beta (z''_{+}+z''_{-})}=1$
and $e^{2\beta z''_{-}}>1 > e^{2\beta z''_{+}}$ and
$e^{2\beta z'_{-}}\le 1 \le e^{2\beta z'_{+}}$
where in the last case, the equality is obtained
when $\beta = 0$-the extreme limit is taken. 
(As one can see from Eq.\ (\ref{gammadet}), there is
no extreme limit ($\beta\rightarrow 0$)\ in the de~Sitter case.)

\subsection{Israel's matching conditions and surface energy density of the domain walls} \label{subsubisrael}

In the thin wall approximation, the energy-momentum tensor and the 
Einstein tensor have $\delta$-function singularities at the wall. 
In Israel's formalism \cite{ISR} for singular layers 
the metric tensor has a discontinuity in its first order
 derivatives in the direction transverse to the wall, and  
the Lanczos tensor ${\cal{S}}^{i}_{\;j}$, which is 
 the surface energy-momentum tensor of the wall located at $z_{0}$,
  is related to the discontinuity of the 
   extrinsic curvature $K^{i}_{\;j}$ in the following way:
\begin{equation}
\kappa_{D} {\cal{S}}^{i}_{\;j}  = - [K^{i}_{\;j}]^{-}  +
\delta^{i}_{\;j}[K]^{-}.
\end{equation}
The square brackets $[\;\;\;]^{-}$ is defined as $[\Omega]^{-} \equiv 
 \lim_{\epsilon\rightarrow 0}(\Omega(z_{0}+\epsilon)) 
 - \Omega(z_{0}-\epsilon))$ and 
$x^{i} \in \{t,r,\phi_i \}$ ($i=1,\cdots , D-3$) are the coordinates
parallel to the wall.
 $K^{i}_{\;j}$ is given by the covariant derivative 
 of the space-like unit normal  $n^{\mu}$ of
  the wall's hyper-space-time.
In a normalized coordinate system where $g_{\hat{z}\hat{z}}=-1$, 
the extrinsic curvature can be written as 
$K_{i j} = -{ \zeta \over 2}g_{ij,\hat{z}}\, $ 
where $\zeta = \pm 1$ signifies 
the inherent sign ambiguity of the unit normal $n^{\mu}$.
Hence, in a  comoving coordinate frame, where we have chosen 
 $A(z_0=0)=1$,  the Lanczos 
tensor can be  written as 
\begin{equation}
\kappa_{D} {\cal{S}}^{i}_{\;j}=-\delta^{i}_{\;j}[\zeta H]^{-}_{z=0}.
\label{sij}
\end{equation}
The  energy density of the wall $\sigma\equiv {\cal{S}}^{t}_{\; t}$, which is equal to the wall's tension $\tau\equiv {\cal{S}}^{r}_{\;r}={\cal{S}}^{\phi_i}_{\;\phi_i}$, is given by
\begin{equation}
\kappa_D\sigma=-[\zeta H]^{-}_{z=0}.
\label{sigdef}
\end{equation}
Applying Israel's formalism to the local
 vacuum solutions  specified by  Eqs. (\ref{smin})-
 (\ref{aplus}), we find the surface energy density $\sigma$, and tension
  $\tau=\sigma$, to be of the form: 
\begin{equation}
\kappa_D\sigma = 2\zeta_{1}h_{1}\left(q_{0}-{\Lambda_{1} \over (D-3)(D-1)}\right)^{1/2}
-2\zeta_{2}h_{2}\left(q_{0}-{\Lambda_{2}\over (D-3)(D-1)}\right)^{1/2} ,
\label{gensig}
\end{equation}
where the first [second] contribution to the energy density  comes
from $z<0$ [$z>0$] side of the wall. We choose, without loss of generality,
to orient the $z$-coordinate so that the
vacuum of lowest energy will be placed
on the $z<0$ side (i.e. $\Lambda_1<\Lambda_2$).

In Eq. (\ref{gensig}), in additional to the ambiguity in the sign of the
unit normal $n^{\mu}$ ($\zeta_{i} =\pm 1$), there is 
another sign ambiguity, $h_{i} = \pm 1$, 
in taking the square-root of Eq.\ (\ref{aeq}). 
A kink-like solution for the wall 
i.e. the scalar interpolating between the extrema of the potential,
 implies   $\zeta_{1} =  \zeta_{2}=1$. 
In addition, we take $h_{i}=1$ if $A_{i}(z)$ is an increasing function of $z$; 
and $h_{i}=-1$ if $A_{i}(z)$ is decreasing.

The domain wall solutions fall into two categories: those with 
 positive energy density (corresponding to 
 the infinitely thin wall limit of a kink solution,
  interpolating between  the  {\it minima} of the potential) 
   and those with
  negative energy density which correspond to choosing 
 the reversed values of $h_i$. 
Examples of negative tension walls are encountered in  gauged supergravity theories (see, e.g., \cite{BC}) as a consequence of a kink solution
 interpolating between {\it maxima} of the potential.

For a given value of $q_0$ the walls can be classified 
 according to the choice of $h_i$  into the following classes 
(the notations are chosen to be  compatible  with
earlier classifications \cite{CGI,BC} of extreme wall ($q_0=0$) solutions):
\begin{itemize}
\item
Type I walls:  a special case  of Type II walls with $\Lambda_2=0$,
and $q_0=0$.
\item
Type II  walls:  positive-tension walls with  $h_1=-h_2=1$.
\item  
Type III  walls: positive-tension walls with $h_1=h_2=1$.
\item
Type III$'$ walls: negative-tension walls with $h_1=h_2=-1$, and
$\sigma_{III'}=-\sigma_{III}$.
\item
Type IV walls: negative-tension walls with $h_1=-h_2=-1$, and $\sigma_{IV}=-\sigma_{II}$. 
\item
Type V walls:  a special case of  of Type IV walls with $\Lambda_2=0$,
$q_0=0$, and  
 $\sigma_{V}=-\sigma_{I}$.
\end{itemize}

The global and local space-times of positive tension walls with $q_0=0$ 
(and $S_0=1$) as well as   $q_0>0$  
 (and $S_+=\cosh(\beta t)$)  were extensively
studied in \cite{CGSII}. In \cite{CGSII}, 
the $q_0<0$  ($S(t)=\cos(\beta t)$) examples were not further 
studied, in part due to the geodesic incompleteness of the space-time
description of the AdS$_2$ space-time transverse to the wall. Nevertheless,
these are proper local solutions  deserving further study (see also 
\cite{170}).
On the other hand  the negative tension walls are of interest  
 in the study of AdS/CFT correspondence and thus deserve further 
investigation. 
 
In the following Section we provide a systematic classification of the
 (local and global) space-time structure of the possible domain wall
 solutions.

\section{Classification of the domain wall solutions}
\label{classification}

We shall  classify the solutions according to the 
 values of the parameter $q_{0}$.
The metric, intrinsic to the wall and specified by $S(t)$
is locally related to standard
coordinates of M$_{D-1}$, AdS$_{D-1}$, or dS$_{D-1}$ space-times
for $q_0=0$, $q_0>0$ or $q_0<0$, respectively.
Within each class we then discuss the space-time  structure transverse 
to the wall as determined by the metric
conformal factor $A(z)$; its structure is governed by the energy density (\ref{gensig}) and
its relationship to the cosmological constants on either side of
the wall. 

\subsection{Walls with $(q_{0}=0)$: extreme walls }

The $q_{0}=0$ solutions, known as extreme domain walls~\cite{CDGS},
exist for $\Lambda_{1,2}\le 0$. 
 (The cosmological constant is defined as 
$\Lambda_{1,2}\equiv  -(D-3)(D-1)\alpha_{1,2}^{2}$.)

Since the wall is homogeneous, isotropic and boost invariant,
the spatial curvature of constant $z$ sections is not
unambiguously defined; 
there is no preferred frame in the (D-1)-dimensional
space-time of the wall. 
The two $S_{0}$ solutions (\ref{snull})---the
Milne type solution with
$S=\beta t$ and $k=-\beta^2$ and the inertial
Minkowski solution with $S=1$ and $k=0$--- both describe
M$_{D-1}$ space-time. The two solutions  are related by a
coordinate transformation \cite{Robertson}  that does not
involve the transverse coordinate $z$ and
therefore describe locally equivalent space-times. 

The $S=1$ solution is the {\em only\/} wall which
represents a noncompact {\em planar\/} ($k=0$) and static wall.
These walls  could be 
realized as supersymmetric bosonic configurations. Examples of 
supersymmetric domain walls in D=4
$N=1$ supergravity coupled to chiral matter superfields were first found and
 studied in~\cite{CGRI},  and recent examples within D=5 
 gauged supergravities  were given in \cite{BC}.

The  physically distinct solutions  (with $S(t)=1$) correspond  to 
 two sets of solutions for $A(z)$ in Eq.\ (\ref{anull}),  with the 
 asymptotic $D$-dimensional anti-deSitter ($AdS_D$) and Minkowski ($M_D$) 
space-times described by 
{\em horo-spherical\/}  and Cartesian 
coordinates, respectively.

Positive energy solutions can be classified
as the following  three types \cite{CGI}, according to the relationship of 
the energy density of the wall $\sigma$  to $\alpha_{1,2}$: 
\begin{itemize}
\item
Type I: 
planar walls with  $\kappa_D\sigma_{ext,I}=2\alpha_1$, 
interpolate between M$_{D}$ and AdS$_{D}$ where on the latter side
 the metric
 conformal factor $A(z)$ decreases  and reaches Cauchy horizon at $z\to
 -\infty$. These walls saturate  a $D$-dimensional analog of the 
Coleman-deLuccia \cite{CD}
 bound.
\item
Type II: planar walls  with  $\kappa_D\sigma_{ext,II}=2(\alpha_1+\alpha_2)$
interpolate between two AdS$_{D}$ regions, in which  $A(z)$
decreases (repulsive gravity)  away from either side of 
the wall. (The special case with a $Z_2$-symmetry ($\alpha_1=\alpha_2$)
 in D=5 gives rise to the Randall-Sundrum scenario with one positive
  tension brane\cite{070}.)    $z=\pm\infty$ correspond to  the Cauchy
AdS horizons.  The geodesic extensions  were studied extensively
in  \cite{CDGS,Gibb,CGSII} and bear striking similarities to the global 
space-times of extreme charged black holes.
\item 
Type III: planar walls  with $\kappa_D\sigma_{ext,III}=2(\alpha_1-\alpha_2)$
interpolate between two AdS$_{D}$ spaces with different
cosmological constants; the conformal
factor goes to infinity on $z>0$ side of the wall, 
while decreases on the other side. 
 The singularity in $A(z)$ at a finite value of $z$  represents the time-like
boundary of the  AdS$_{D}$ space-time. Again $z\to-\infty$  corresponds
to the Cauchy horizon.
\end{itemize}

\begin{figure}
\centerline{
\hbox{
\epsfxsize=1.7truein
\epsfbox[70 32 545 740]{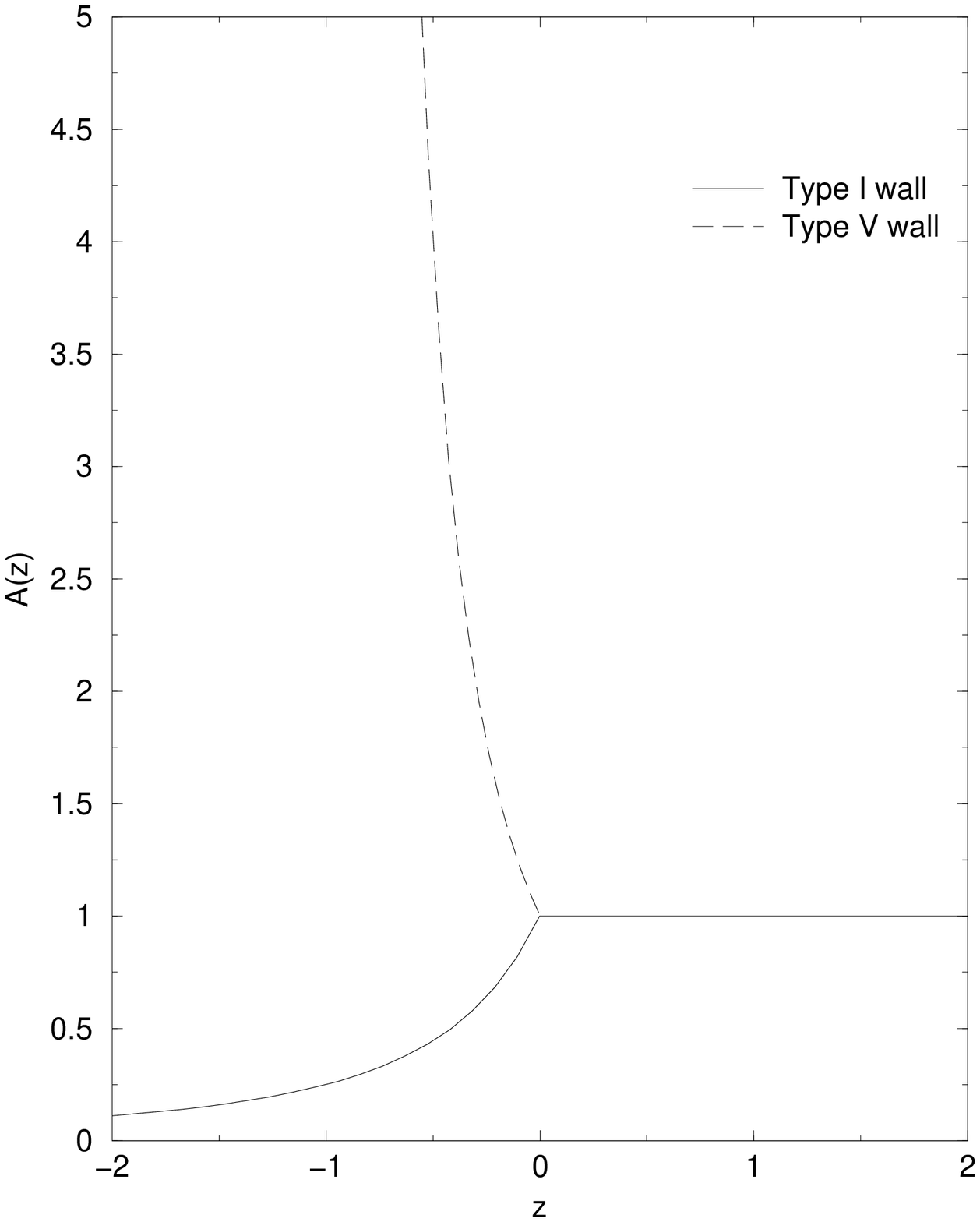}
\hskip 0.25truein
\epsfxsize=1.7truein
\epsfbox[70 32 545 740]{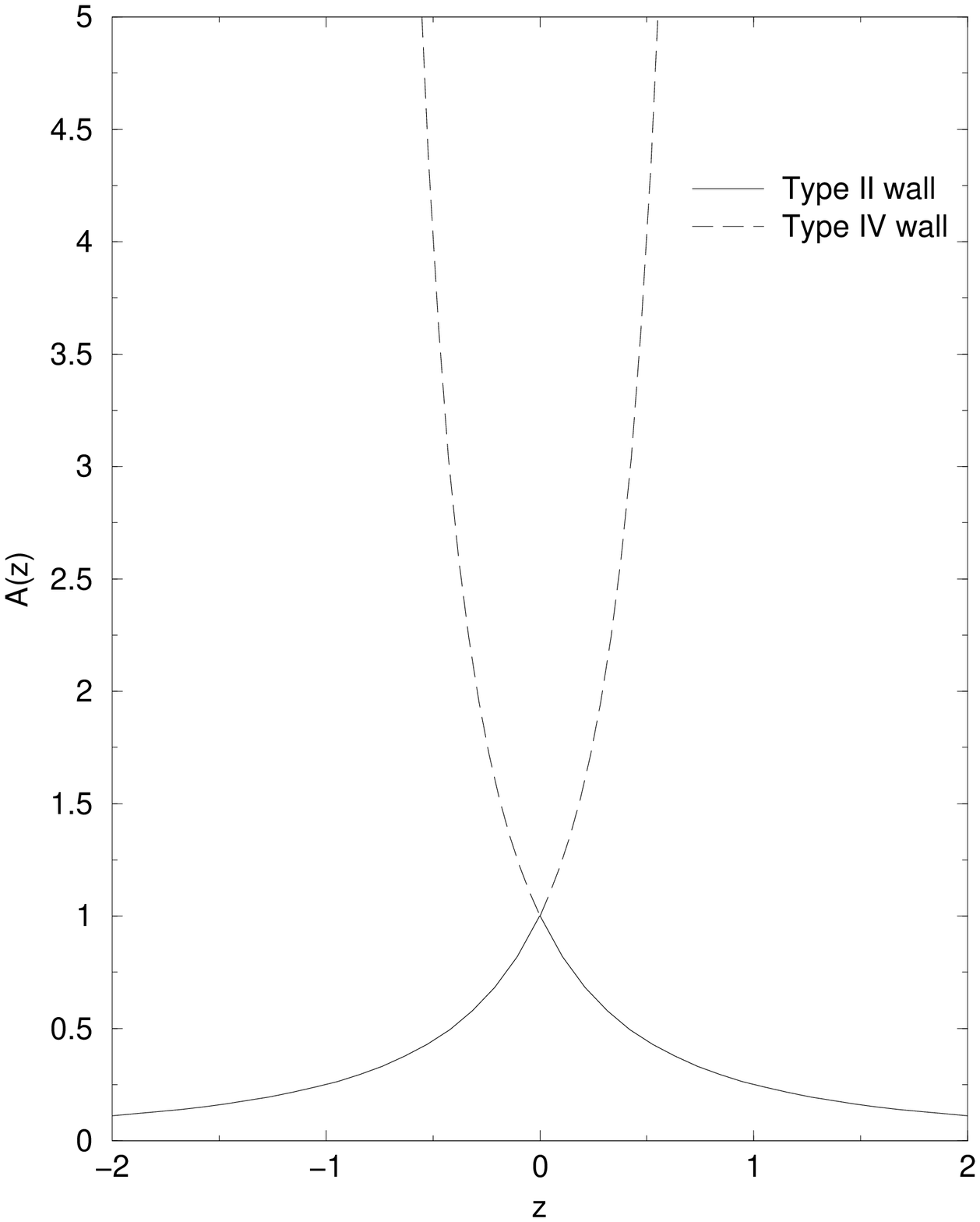}
\hskip 0.25truein
\epsfxsize=1.7truein
\epsfbox[70 32 545 740]{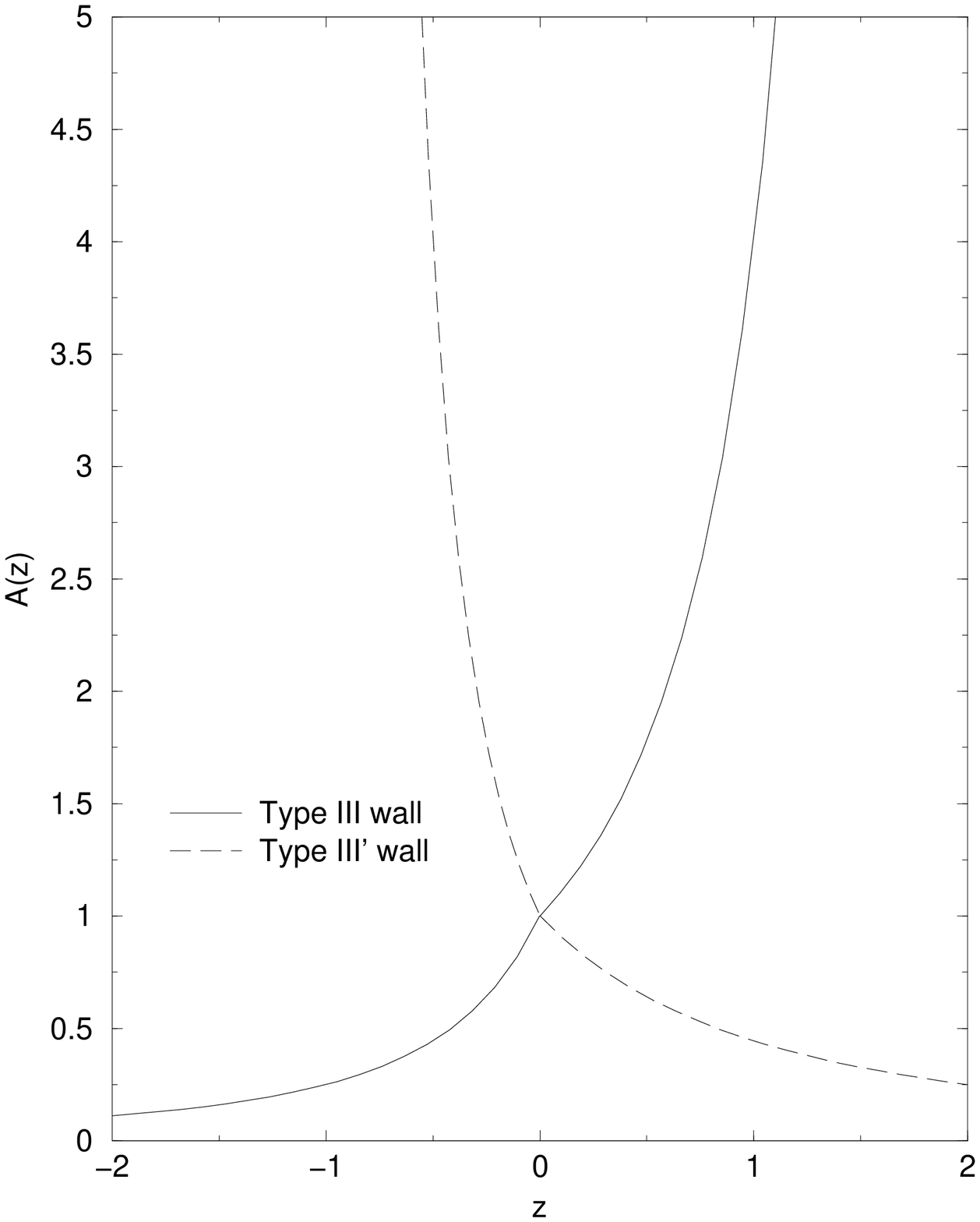}
}
}
\caption{
The  metric coefficients $A(z)$  are plotted 
as a function of transverse coordinate $z$, 
for  dual pairs of extreme ($q_0=0$) domain  walls.
  Fig.(1a) denotes  Type I-Type V pairs with the cosmological constant
  parameters $\alpha_1=1$ and $\alpha_2=0$,
  Fig.(1b) shows Type II-Type IV pairs with  
$\alpha_1=\alpha_2=1$ and Fig.(1c) shows Type III-Type III$'$ pairs with
$\alpha_1=1$ and $\alpha_2 = 1/2$. }
\end{figure} 
 
On the other hand, the walls with negative energy density fall into the
following classes \cite{BC}:
\begin{itemize}
\item 
Type III$'$: planar walls  with $\sigma_{ext,III'}=-2(\alpha_1-\alpha_2)$
 have the space-time structure that is a mirror image of that of 
 Type III walls; the conformal factor goes to
  infinity on $z<0$ side of the wall,
on $z>0$ side the conformal factor decreases,  
  the AdS Cauchy horizon is at $z\to \infty$.
\item
Type IV: planar walls with $\sigma_{ext,IV}=-2(\alpha_1+\alpha_2)$ 
 interpolate between two
AdS$_{D}$ regions, in which  $A(z)$ increases on either side 
of the wall,
reaching the  AdS  boundaries at a finite value of $z$ on 
either side. (Those are
typical wall solutions encountered in gauged supergravity 
theories (see, e.g., \cite{BC}), and are of interest in the study of RGE
flows in the context of AdS/CFT correspondence.) 
 \item
Type V: planar walls with  $\sigma_{ext,V}=-2\alpha_1$,
interpolating between M$_{D}$ and AdS$_{D}$; on the AdS side 
$A(z)$ increases away from the wall, approaching the
boundary of the AdS at a finite value of $z$.

\end{itemize}

The behavior of the conformal factor $A(z)$ for each class of solutions 
can be easily seen from specific examples shown in Fig.1.  
 Type I-V, Type II-IV and Type III-III$'$ can be viewed as ``dual''. 
Namely,  the 
  energy density of these walls have opposite signs  and the 
   space-time patches of the  AdS$_{D}$  are complementary, i.e.,  AdS Cauchy
   horizons in one case are replaced by the boundaries of AdS  space-times in another and  vice  versa. 
 Note  that in this sense the  extreme  Type II walls, which
  provide a realization of
the Randall-Sundrum scenario in D=5,  and Type IV  walls,  which are
generically encountered in  the study
of AdS/CFT correspondence, provide the {\it complementary} domains of the 
AdS space-time.  We also note that within field-theoretic framework,
 such as gauged supergravity
theories, a  realization of  (finite) negative tension domain walls and 
the  issues of their stability require further study.

\subsection{Walls  with $q_{0}>0$}

$q_{0}=\beta^2$ solutions with  
 the form of $S_{+}$ in Eq. (\ref{splus}) describe 
(D-1)-dimensional de~Sitter space-time (dS$_{D-1}$) parallel to the wall.
The topology of dS$_{D-1}$ is 
${\bf R}({\text{time}})\times {\bf S}^{D-2}({\text{space}})$.
dS$_{D-1}$ represents a hyperboloid embedded in a flat D-dimensional Minkowski space-time \cite{Gibb},
and the three possible spatial curvatures $(k=-\beta^2,~0,~\beta^2)$ 
correspond to three different choices of constant time slices of this
hyperboloid. However, only the positive curvature solution with $S(t)=\cosh
(\beta t)$  yields the complete covering of dS$_{D-1}$ \cite{Gibb,CGSII}; we 
will mainly focus on this class of solutions, which  have a topology of the 
 {\it ``expanding bubbles''}. A possible way to   create 
such expanding bubbles is via instantons of  Euclidean  gravity.


The walls can be classified according to their energy density (\ref{gensig})
into the following classes:
\begin{itemize}
\item
Type II walls with
\begin{equation}
\kappa_D\sigma_{{non, II}}=2\left(\pm\alpha^2_{1}+\beta^2\right)^{1/2}+
2\left(\pm\alpha^2_{2}+\beta^2\right)^{1/2}\ge \kappa_D \sigma_{ext,II}.
\label{nex}
\end{equation}
are non-extreme, i.e. their energy density is above that of their extreme
counterparts.  Here 
$\Lambda_{i}\equiv \mp(D-3)(D-1)\alpha_{i}^2$.  Note that the domain walls in
this class involve positive, zero and negative cosmological constants, 
and the minus sign in front of $\alpha_i^2$ corresponds to a 
{\em positive\/} $\Lambda_i$. Further more, in the de~Sitter case,
$\alpha^{2}_{i} \leq \beta^{2}$ is required. There are $6$ possible 
configurations of Type II wall interpolating between different 
space-times, which are shown as $6$ examples in Fig. 2.

Configurations with geodesically complete space-time internal to the wall
($S(t)=\cosh(\beta t)$) describe {\it expanding bubbles with two
insides}.\cite{CGSII}. Namely, because 
 the radius  $R_b$ of the curvature of concentric shells at distance $z$ 
is  proportional to $A^{1/2}(z)S(t)$, $R_b$, at fixed $t$,
  decreases as $A(z)$ decreases with increasing $|z|$,  and therefore 
  either side of the wall corresponds to an {\it inside} region of the bubble.
 In addition, 
  since $S(t)$ increases with $t$, the bubble is  {\it expanding} 
   to an asymptotic observer on either side of the wall. 
  (One possible origin of a creation of such configurations
   is via instantons of Euclidean gravity-quantum
cosmology \cite{QuantumCosmology}).
 
 Since $A(z)$  decreases with increasing $|z|$, gravity is repulsive 
 on either side of the wall and    $z \rightarrow \pm \infty$ corresponds to
 (cosmological) horizons.  The geodesic extensions are  studied  in
 \cite{CGSII}, and bear striking similarities to non-extreme charged
black holes where time-like singularities are replaced by wall
boundaries.

The walls with $\Lambda_{1,2}\le 0$ are generalizations
 of the extreme Type II and Type I walls with $\beta=0$ to $\beta >0$.  
  The walls with $\Lambda_{1}$ or $\Lambda_{2}>0$ and 
$\Lambda_1=\Lambda_2=0$ do not have
an   extreme limit ($\beta\to 0$). 
The latter class of  $M_D-M_D$ Type II domain walls in D=4 
(and $S(t)=e^{\beta t}$) was   studied  in \cite{Vilenkin}.
 dS-dS Type II walls  are unstable, since false vacuum decay walls (dS-dS Type
 III walls)  are dynamically preferred \cite{SATO,BKT}, except for the case of
 $\Lambda_1=\Lambda_2>0$.

\item 
Type III walls with $q_0>0$ have  an energy density lower than that 
of their  extreme wall counterparts: 
\begin{equation}
\kappa_D\sigma_{{ultra,III}}=2\left(\pm\alpha^2_{1}+\beta^2\right)^{1/2}-
2\left(\pm\alpha^2_{2}+\beta^2\right)^{1/2}\le \kappa_D \sigma_{ext,III},
\label{uex}
\end{equation}
and are referred to as ultra-extreme domain walls. 
The solutions with asymptotic $dS_{D}$ space-times require  
$\alpha^{2}_{i} \leq \beta^{2}$, and consequently 
they  do not have an extreme limit.
Specific examples of $5$ possible configurations of the Type III 
walls are shown in Fig.3.

Configurations (with $S(t)=\cosh(\beta t)$)
are {\it false vacuum decay bubbles} \cite{Coleman,BKT}.   Namely, 
the radius  $R_b \propto A^{1/2}(z)S(t)$ decreases  (increases) for $z<0$
($z>0$), and thus   corresponds to the  {\it inside}  ({\it outside}) region
the bubble.  The solutions which only involve M$_{D}$ or AdS$_{D}$, are 
more like ordinary bubbles compared with the Type II walls because   
as $t$ increases the expanding bubble eventually sweeps 
out the space-time on the $z>0$ side. 

On the other hand, on the de Sitter side of the wall, 
the metric function turns around at point $z_{crit}$ and decreases
beyond $z_{crit}$. Hence, beyond $z_{crit}$, the inside of the bubble 
becomes an outside. $z\to \infty$ corresponds to cosmological horizons that 
can be reached by  test particles with energy larger than E$_{crit}$.
In D=4 the non-negative cosmological constant domain walls of this type
were extensively studied 
  in \cite{SATO,BKT}.

  $D=4$ false vacuum decay bubbles 
with non-positive cosmological constants were studied
in \cite{CD,CGSII}.
The inside of the bubble ($z\le 0$)  has the same space-time structure (with
cosmological horizons at $z\to -\infty$),  just as  the Type
II  non-extreme walls ($q_0>0$).  The outside ($z\ge 0$) of the bubble 
 has no horizons and on the M$_{D}$ side of the wall $z\to \infty$ 
corresponds to the boundary of the space-time, while on the AdS side the 
affine boundary is at some finite value of $z$.  

\item Type III$'$  have the energy density:
$\sigma_{non,III'}=-\sigma_{ultra,III}\ge \sigma_{III,ext}$, which is above 
 the corresponding  extreme ($q_0=0$) counterparts.  Their space-time
is also {\it complementary} to that of the  Type III walls (see Fig.3). 
These are ``false
vacuum decay bubbles'' with the {\it larger} cosmological
constant side ($z\ge 0$) sweeping out the vacuum with the smaller
cosmological  constant ($z\le 0$), in most of the cases, except the dS-dS 
wall as shown in Fig.(3e). In the latter case, the metric function $A(z)$ 
becomes a decreasing function of $z$ for $z<z_{crit}$ such that the inside of
 the bubble becomes an outside, and  $z\to
-\infty$ is a cosmological horizon.   

These configurations resemble the dynamics of 
 an ``up-side down world'' and an actual realization of
such  negative tension configurations  within a field
 theoretical framework is needed.
   
\item Type IV walls with $\sigma_{ultra,IV}=-\sigma_{non,II}
\le \sigma_{ext,IV}$ 
are ultra-extreme negative tension   ``{\it expanding bubbles with two 
outsides}''. For non-positive cosmological constants
this is the  ``apocalypse world'' where on either side of the wall an
asymptotic observer will be eventually  hit by  the bubble. 
Namely,   the     
conformal  factor $A(z)$ increases on either side of the wall
reaching the boundary of the space-time, which is $z= \pm \infty$ for  
M$_{D}$ or a finite value of $z$ for AdS$_{D}$,  thus either side 
corresponds to the outside of the wall. Since  $S(t)$  grows 
with $t$, these are expanding bubbles which always hit an asymptotic
observer.

\end{itemize}

\begin{figure}
\centerline{
\hbox{
\epsfxsize=1.7truein
\epsfbox[70 32 545 740]{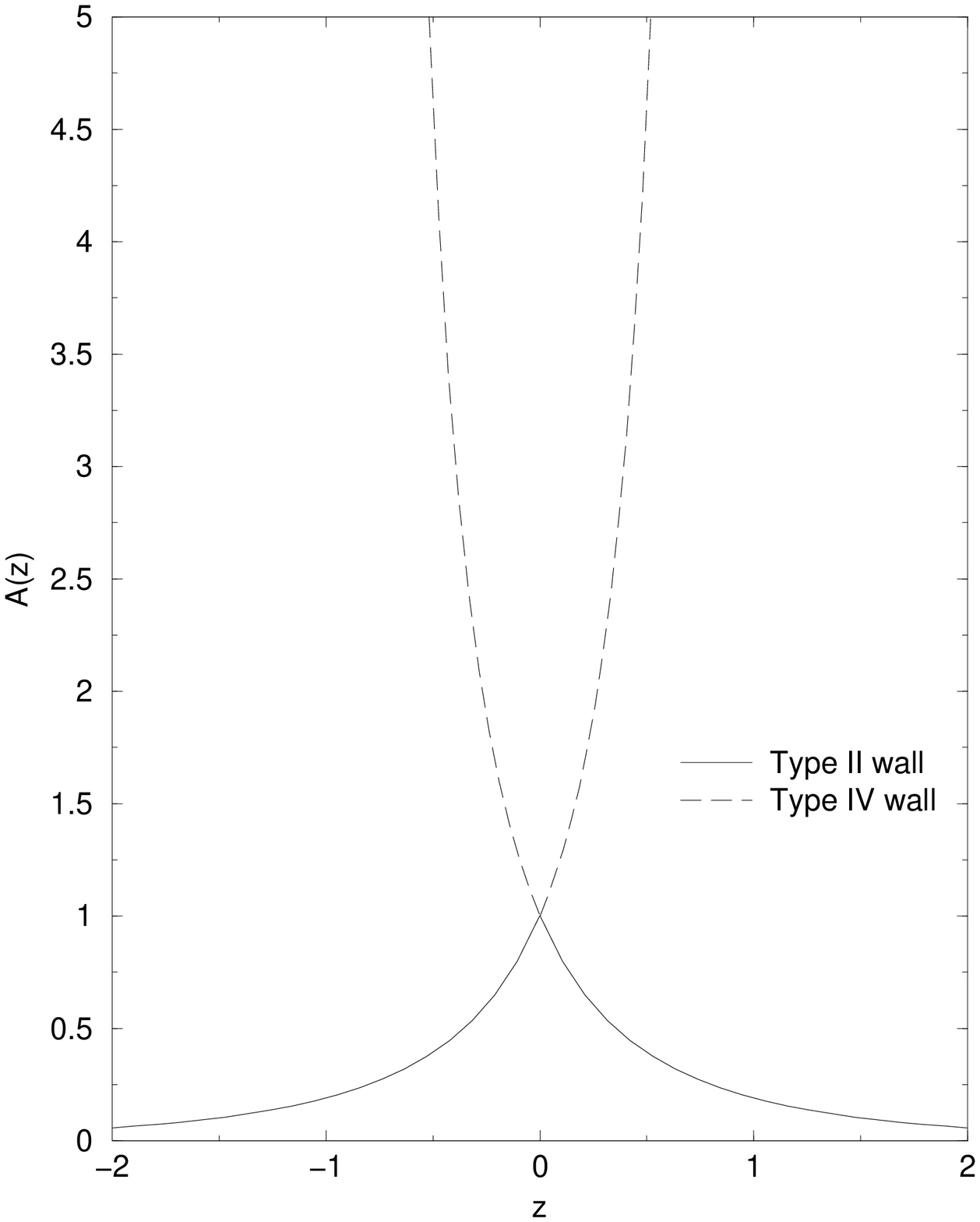}
\hskip 0.25truein
\epsfxsize=1.7truein
\epsfbox[70 32 545 740]{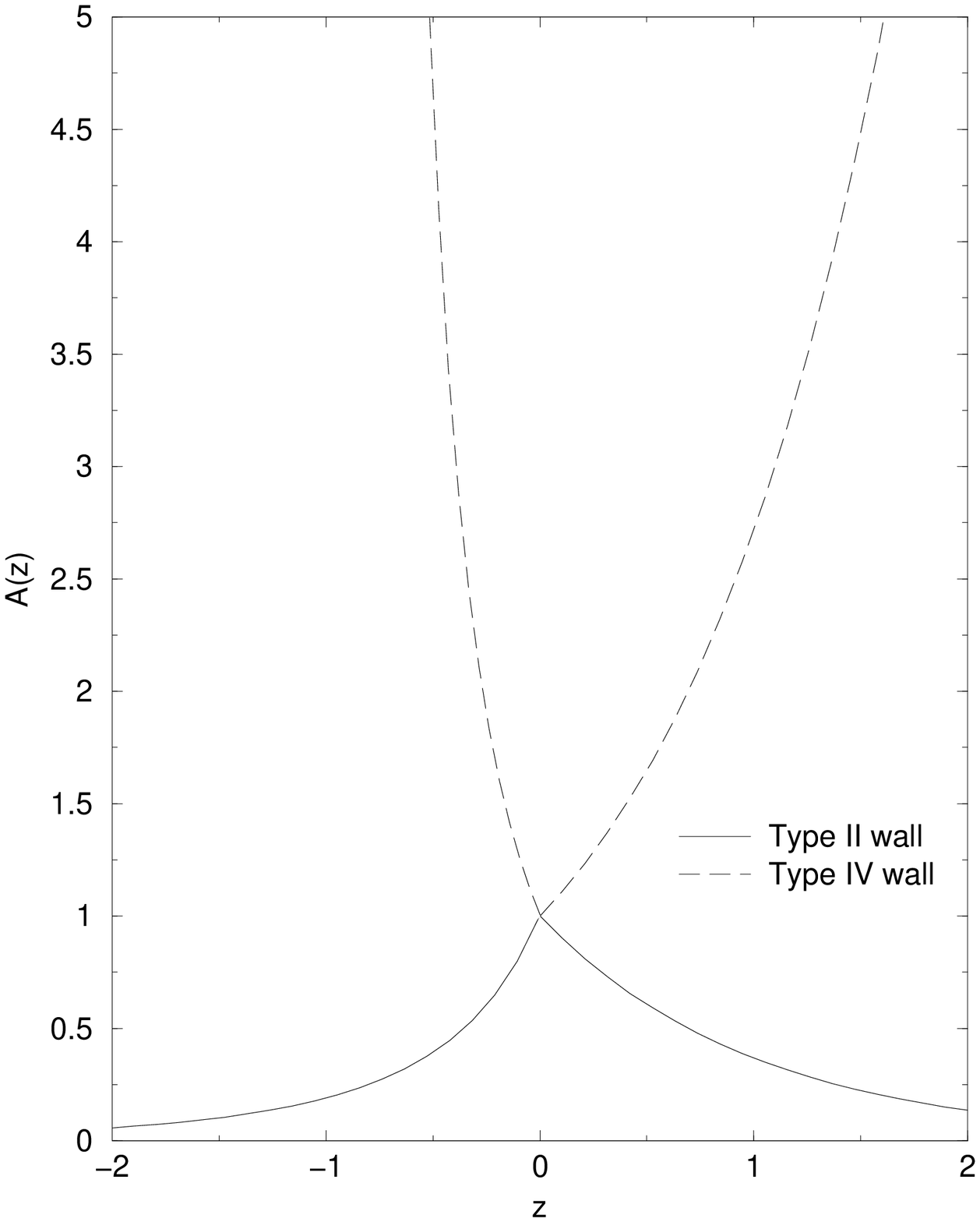}
}
}
\vskip 0.1truein
\centerline{
\hbox{
\epsfxsize=1.7truein
\epsfbox[70 32 545 740]{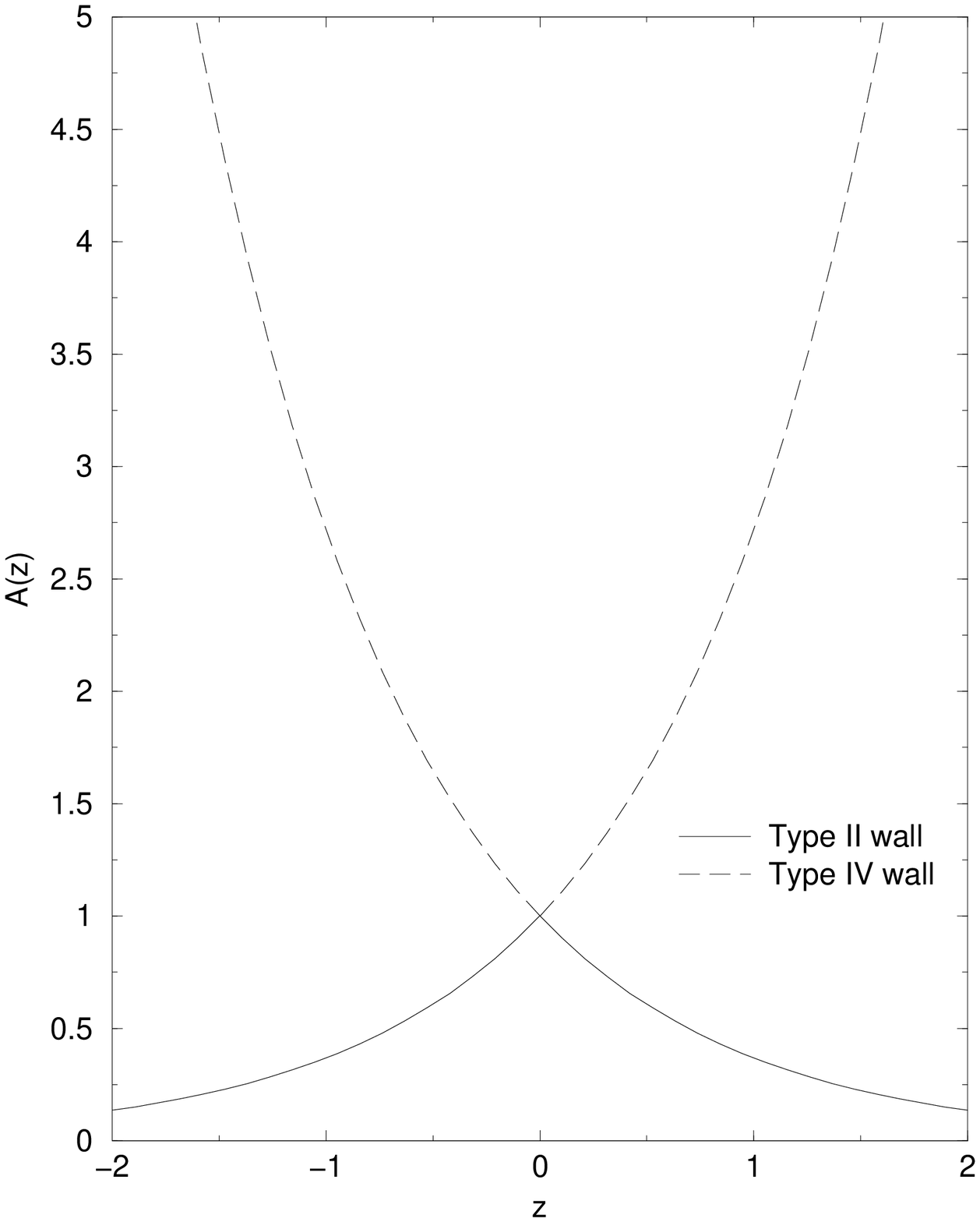}
\hskip 0.25truein
\epsfxsize=1.7truein
\epsfbox[70 32 545 740]{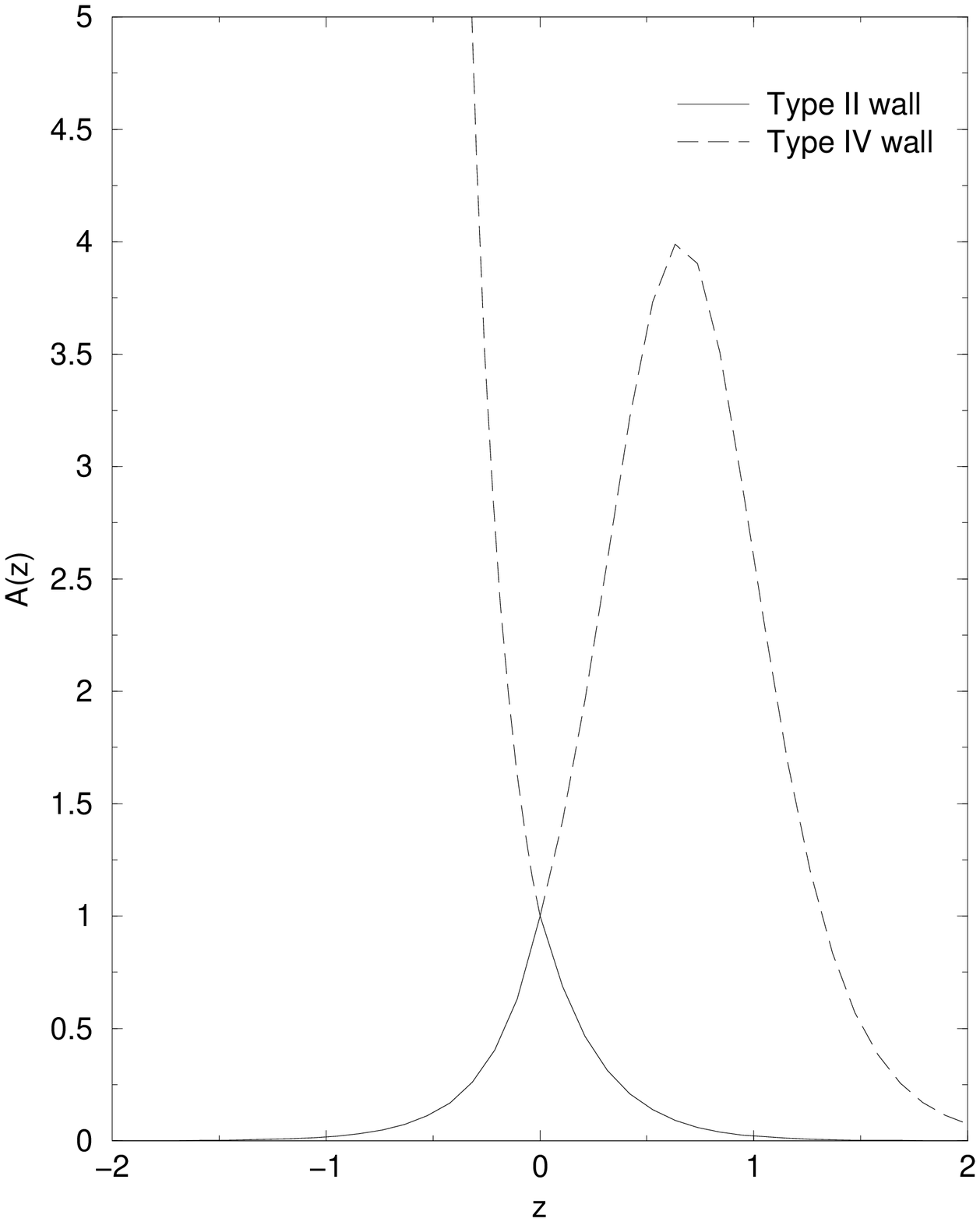}
}
}
\vskip 0.1truein
\centerline{
\hbox{
\epsfxsize=1.7truein
\epsfbox[70 32 545 740]{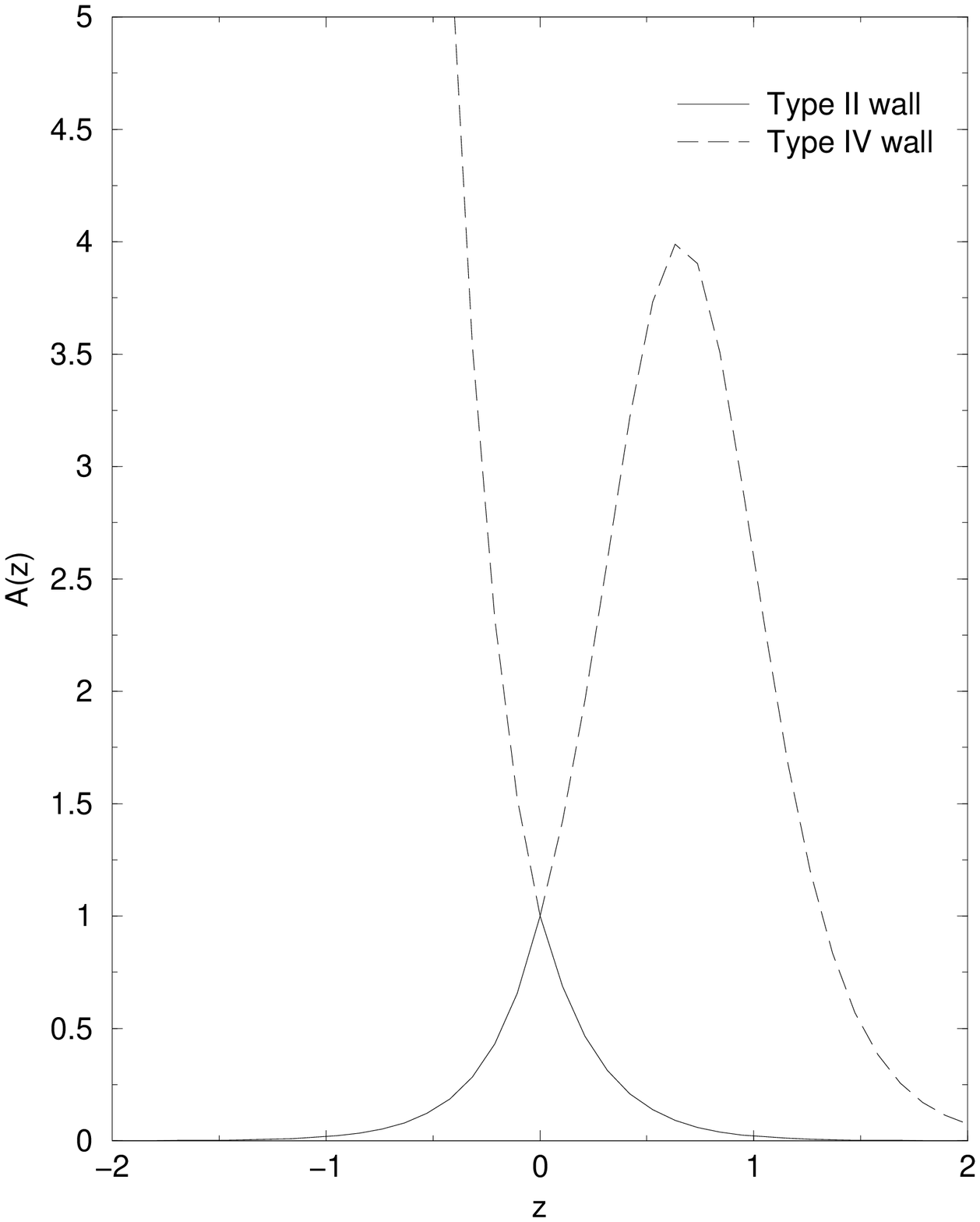}
\hskip 0.25truein
\epsfxsize=1.7truein
\epsfbox[70 32 545 740]{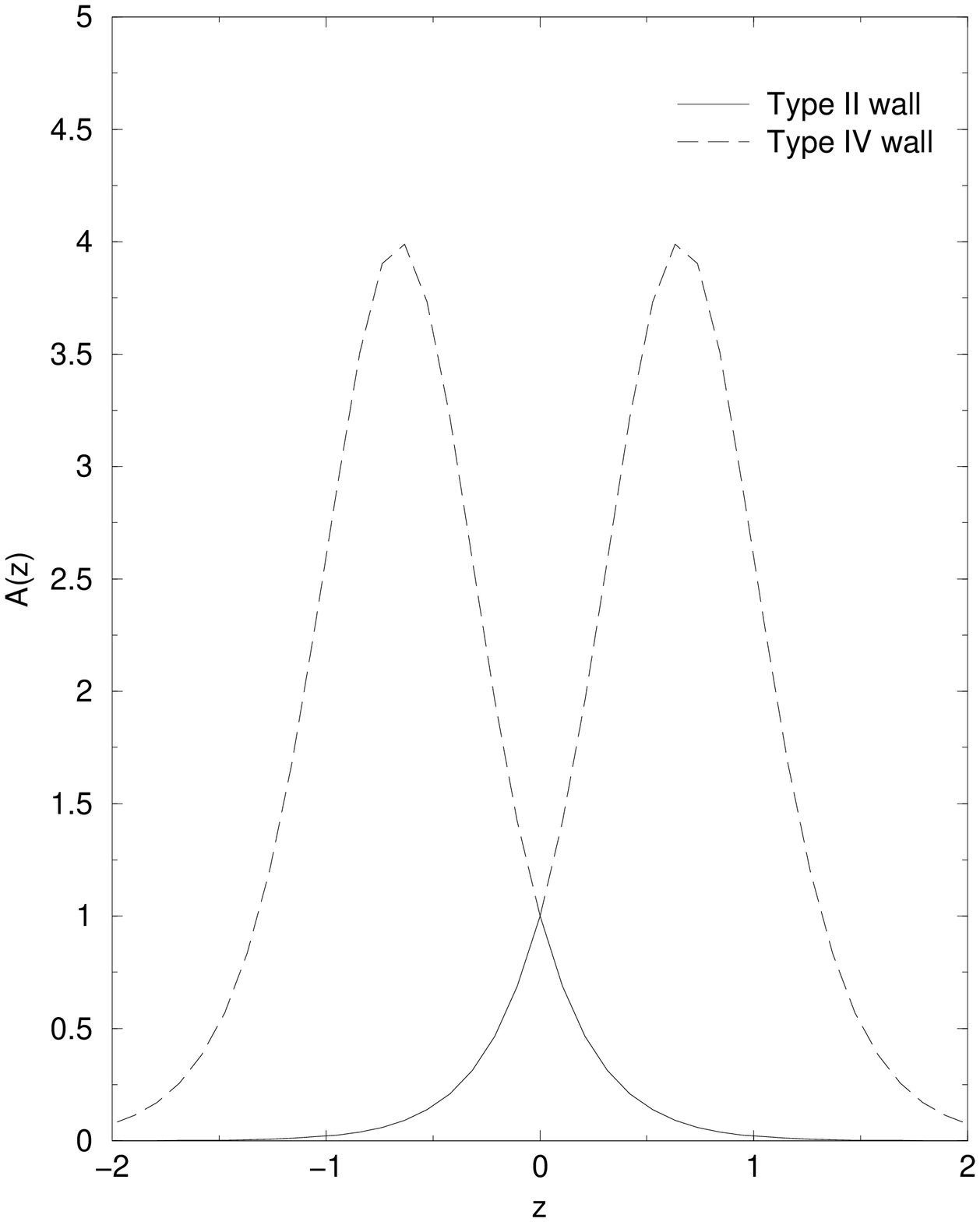}
}
}
\caption{
The metric function $A(z)$ of Type II and Type IV walls in the case of ($q_0>0$). There are six configurations. Fig.(2a) represents a AdS-AdS wall with $\alpha_1= \alpha_2 =1$ and $\beta=1/2$, Fig.(2b) represents a AdS-M wall with $ \alpha_1=1,~ \alpha_2 =0$ and $\beta=1/2$. Fig.(2c) represents a M-M wall with $\beta=1/2$, Fig.(2d) represents a AdS-dS wall with $ \alpha_1= \alpha_2 =1$ and $\beta=2$, Fig.(2e) is  a M-dS wall with $ \alpha_1=0,~ \alpha_2 =1$ and $\beta=2$ and Fig.(2f) represents a dS-dS wall with $ \alpha_1= \alpha_2 =1$ and $\beta=2$. 
}
\end{figure} 

\begin{figure}
\centerline{
\hbox{
\epsfxsize=1.7truein
\epsfbox[70 32 545 740]{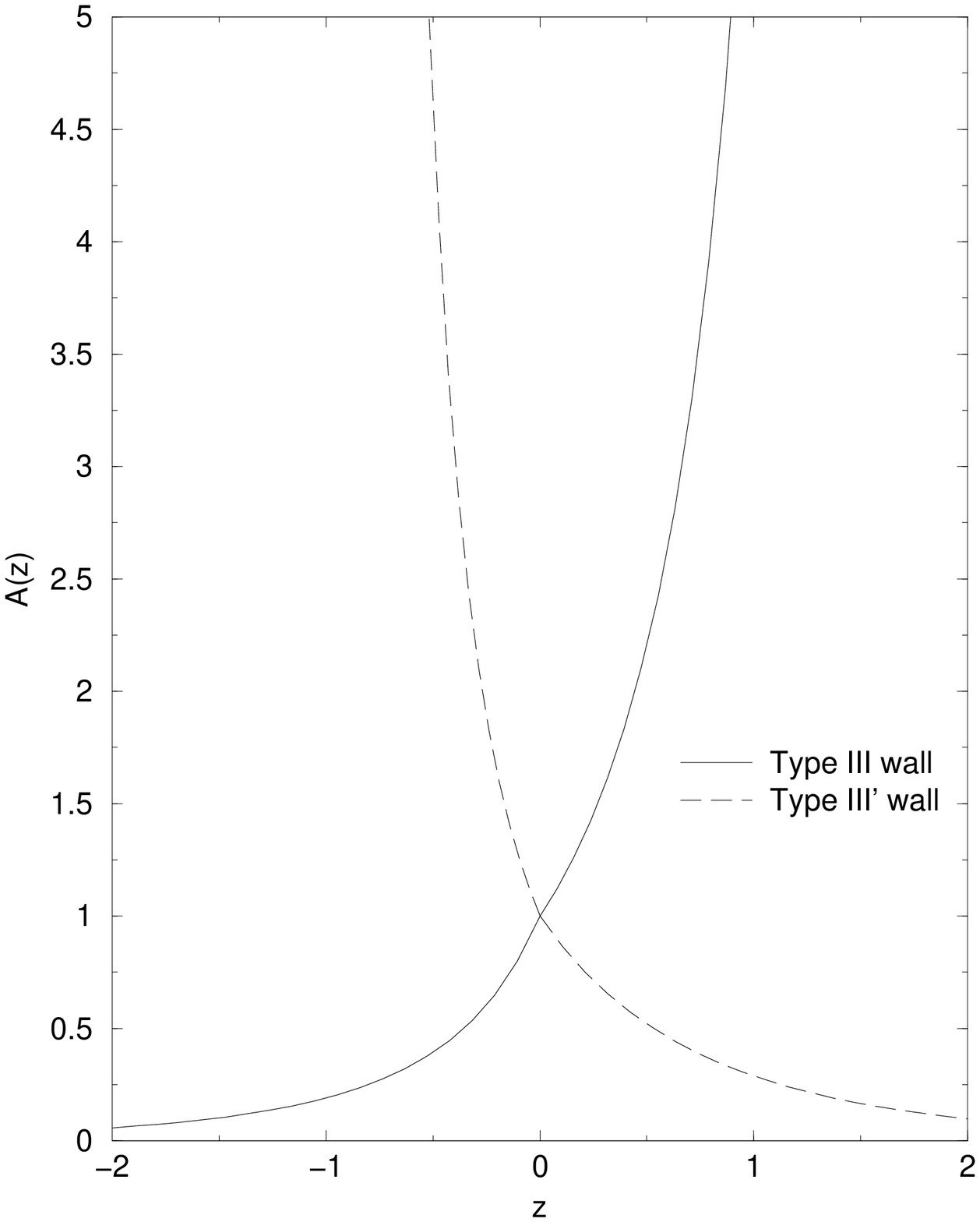}
\hskip 0.25truein
\epsfxsize=1.7truein
\epsfbox[70 32 545 740]{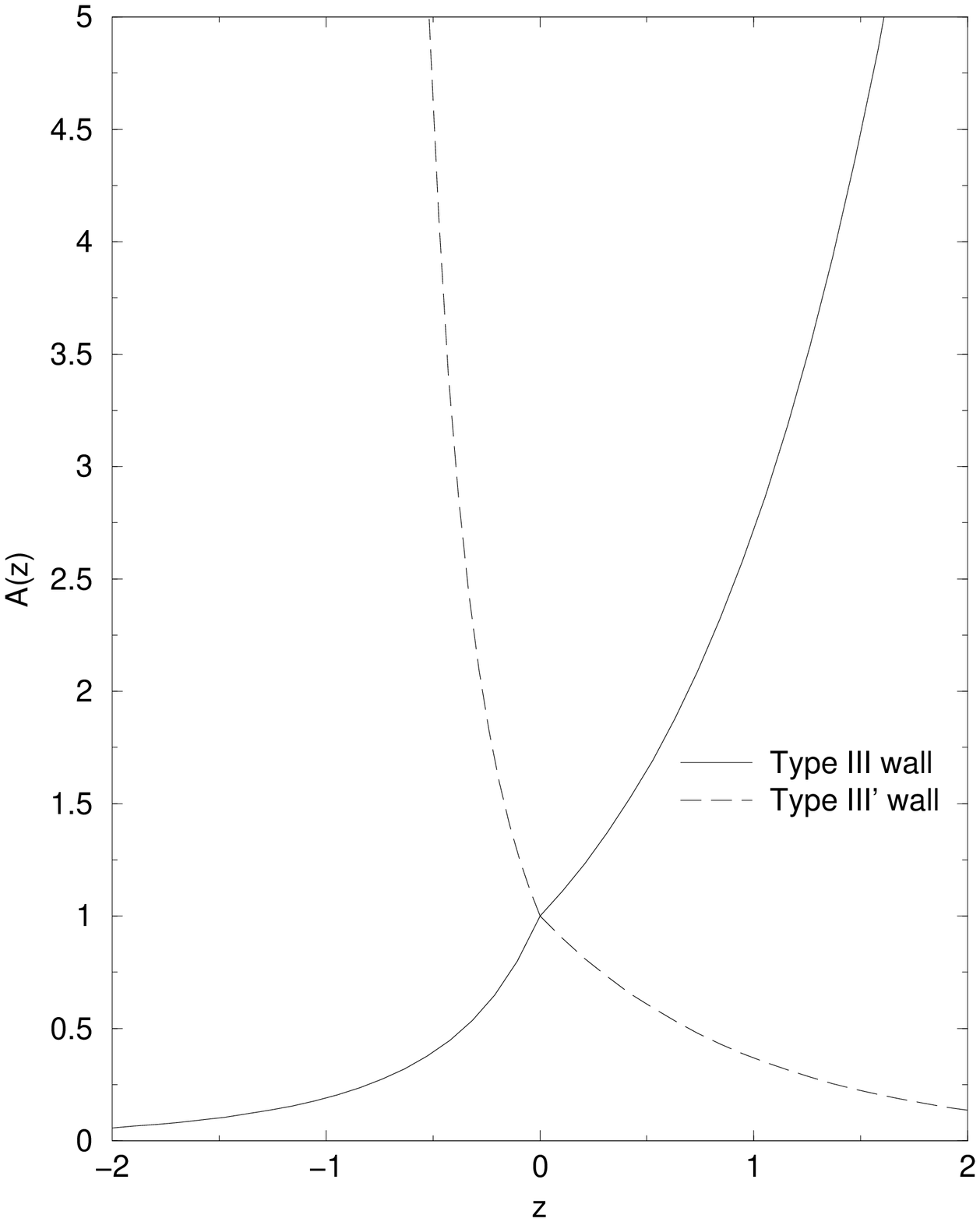}
}
}
\vskip 0.1truein
\centerline{
\hbox{
\epsfxsize=1.7truein
\epsfbox[70 32 545 740]{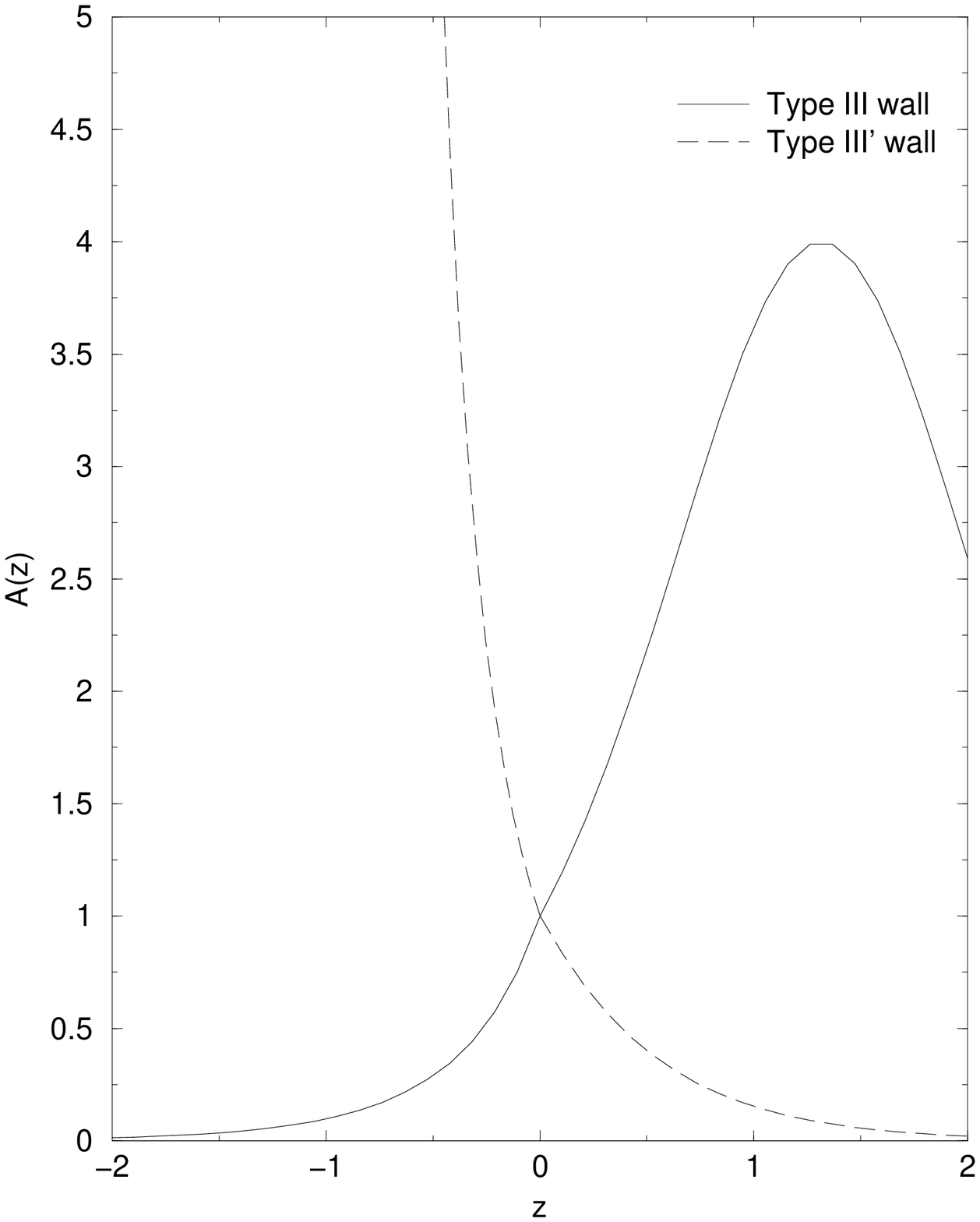}
}
}
\vskip 0.1truein
\centerline{
\hbox{
\epsfxsize=1.7truein
\epsfbox[70 32 545 740]{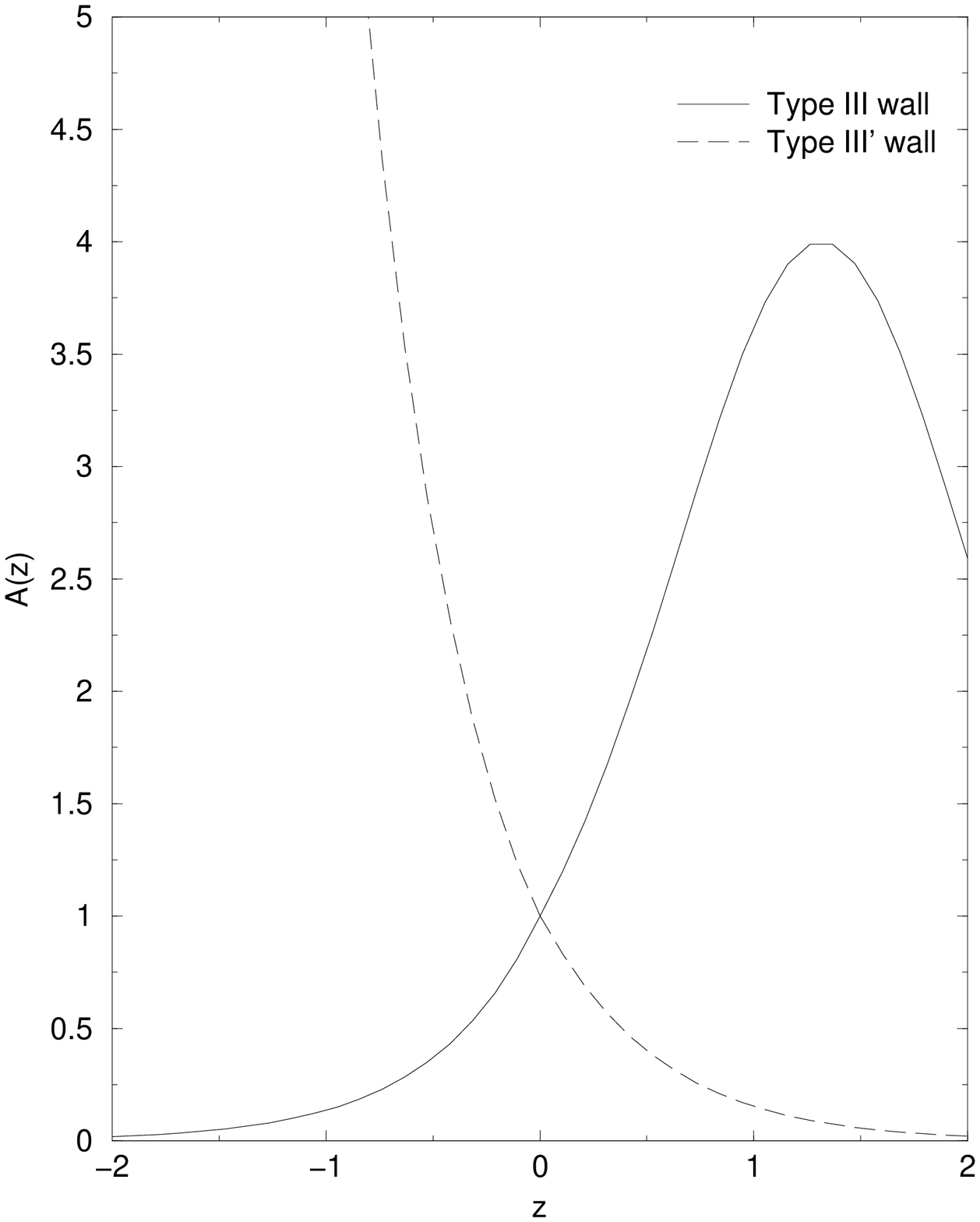}
\hskip 0.25truein
\epsfxsize=1.7truein
\epsfbox[70 32 545 740]{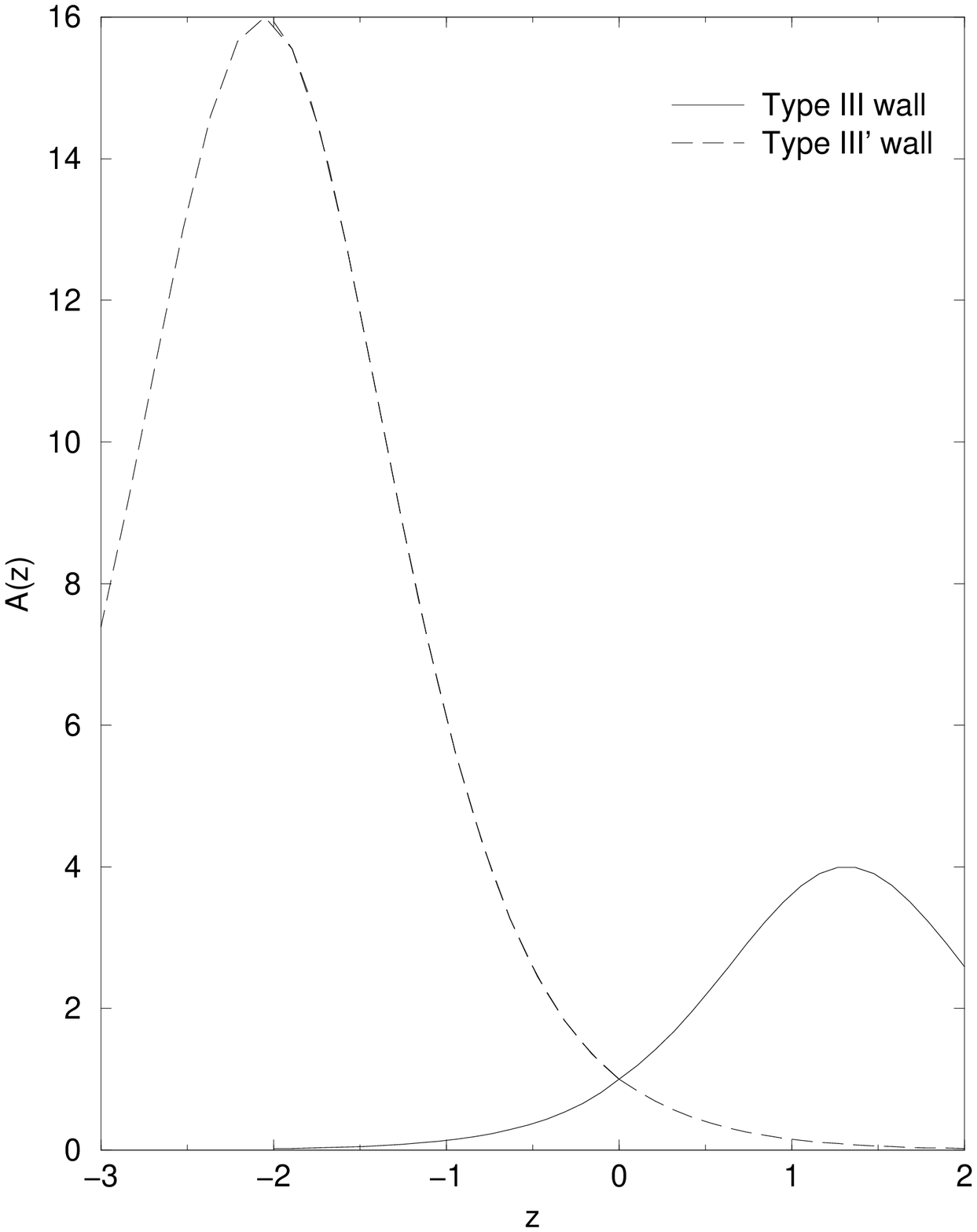}
}
}
\caption{
The metric function $A(z)$ of Type III and Type III' walls in the case of ($q_0>0$). There are five configurations. Fig.(3a) represents a AdS-AdS wall with $\alpha_1=1~, \alpha_2 =1/2$ and $\beta=1/2$, Fig.(3b) represents a AdS-M wall with $ \alpha_1=1,~ \alpha_2 =0$ and $\beta=1/2$, Fig.(3c) represents a AdS-dS wall with $ \alpha_1=1,~ \alpha_2 =1/2$ and $\beta=1$, Fig.(3d) is  a M-dS wall with $ \alpha_1=0,~ \alpha_2 =1/2$ and $\beta=1$ and Fig.(3e) represents a dS-dS wall with $ \alpha_1=1/4,~ \alpha_2 =1/2$ and $\beta=1$. 
}
\end{figure} 
\subsection{Walls with $q_{0}<0$}

For  the case $q_{0} = -\beta^{2}$ the solutions exist only for the 
negative-cosmological constant vacua  that satisfy
$\alpha_i^2={ {-\Lambda_i}\over {(D-1)(D-3)}}\ge \beta^2$,
The space-time internal to the wall  is 
described by a  unique solution with Eq.(\ref{smin}) and Eq.(\ref{amin}); 
 the  space-time intrinsic to the wall  is AdS$_{D-1}$.
 Note that with $S_{-}(t)=\cos{(\beta t)}$ (Eq.(\ref{smin})), 
the region $-{\pi\over 2 \beta}\le t \le 
{\pi\over 2 \beta}$ describes  only a patch of
 the AdS$_{D-1}$ space-time, and  $t=\pm{\pi\over 2 \beta}$ 
 corresponds to an
 apparent coordinate singularity (see e.g., \cite{HE}).
Also, since the radius of the curvature transverse to the wall
$R_b\propto |S(t)|$, the wall is expanding for $-{\pi\over {2\beta}}\le t\le 0$ and then
shrinking for $0\le  t\le {\pi\over {2\beta}}$.

These walls   could be viewed as a generalization of their 
 extreme  counterparts (with $q_0=\beta=0$) to $q_{0}<0$. They can
also be viewed as   
 an analytic continuation of AdS-AdS domain walls with $q_0= \beta ^2>0$ to 
imaginary $\beta$. In this sense the AdS-AdS 
walls with $q_0>0$ and $q_0<0$ are ``dual'' to each other, 
and extreme AdS-AdS walls with $q_0=0$
provide a dividing line between the two classes of solutions.

The walls with $q_0<0$ can be classified  according to their energy 
 densities relative to their extreme counterparts as:
\begin{itemize}
\item 
Type II  walls  are ultra-extreme walls with $
\kappa_D \sigma_{ultra,II} = 
 2( \alpha_{1}^2 - \beta^2)^{1/2} + 2( \alpha_{2}^2 -
 \beta^2)^{1/2}\le \kappa_{D}\sigma_{ext,II}$.  
The space-time transverse to either side of
 the wall has repulsive gravity near the wall,
i.e. $A(z)$ decreases away from the wall until 
the critical point $\beta z_{crit}+ \vartheta_{\pm} \equiv 0$. 
Beyond $z_{crit}$, $A(z)$ increases with increasing $|z|$, 
reaching the affine boundary of the AdS space at 
$\beta z + \vartheta_{\pm} =\frac{\pi}{2}$. 
Therefore, these walls exhibit  repulsive gravity  only
 in the region close to the wall. 
Eventually, geodesics
cross into  a region of attractive gravity, with only null
geodesics reaching the AdS boundary.  Interestingly, there are 
no cosmological horizons. (Note the conformal factor $A(z)$ has  a
complementary behavior relative to that of $q_0>0$ dS-dS wall.)

\item 
Type III walls  are non-extreme  walls with
$\kappa_D \sigma_{non,III} = 2( \alpha_{1}^2 - \beta^2)^{1/2} - 
2( \alpha_{2}^2 - \beta^2)^{1/2}\ge \sigma_{ext,III}.$ 
On the $z<0$ side the gravity is again first attractive and then repulsive, 
 the point $\beta z +\vartheta_{+}=-{\pi\over 2}$ corresponding 
to the AdS boundary. On the
other hand, on the $z>0$ side of the wall, gravity is attractive 
   with  the AdS  boundary taking place at $\beta z+\vartheta_{-}={\pi\over 2}$. 
\item
Type III$'$ walls are ultra-extreme walls  with 
  $\kappa_D \sigma_{ultra,III'} = -2( \alpha_{1}^2 - \beta^2)^{1/2} + 
2( \alpha_{2}^2 - \beta^2)^{1/2}\le \kappa_{D} \sigma_{ext,III'}$.  
Its space-time structure in the transverse direction 
is a mirror image of the Type III $q_{0}<0$ walls. 

\item 

Type IV walls are ultra-extreme walls with 
 $\kappa_D \sigma_{non,IV} =
 -2( \alpha_{1}^2 - \beta^2)^{1/2} - 2( \alpha_{2}^2 -
 \beta^2)^{1/2}\ge \kappa_{D} \sigma_{ext,IV}$, with 
 attractive gravity  on either-side of the wall until 
   $\beta z+\vartheta_{\pm}= \pm {\pi \over 2}$, the 
   the AdS boundary. 
\end{itemize}

 It would be very interesting to investigate further the global space-time  properties of
 these configurations, the issues of their dynamic stability,  as well
  as their field theoretic  embedding.

\begin{figure}
\centerline{
\hbox{
\epsfxsize=1.7truein
\epsfbox[70 32 545 740]{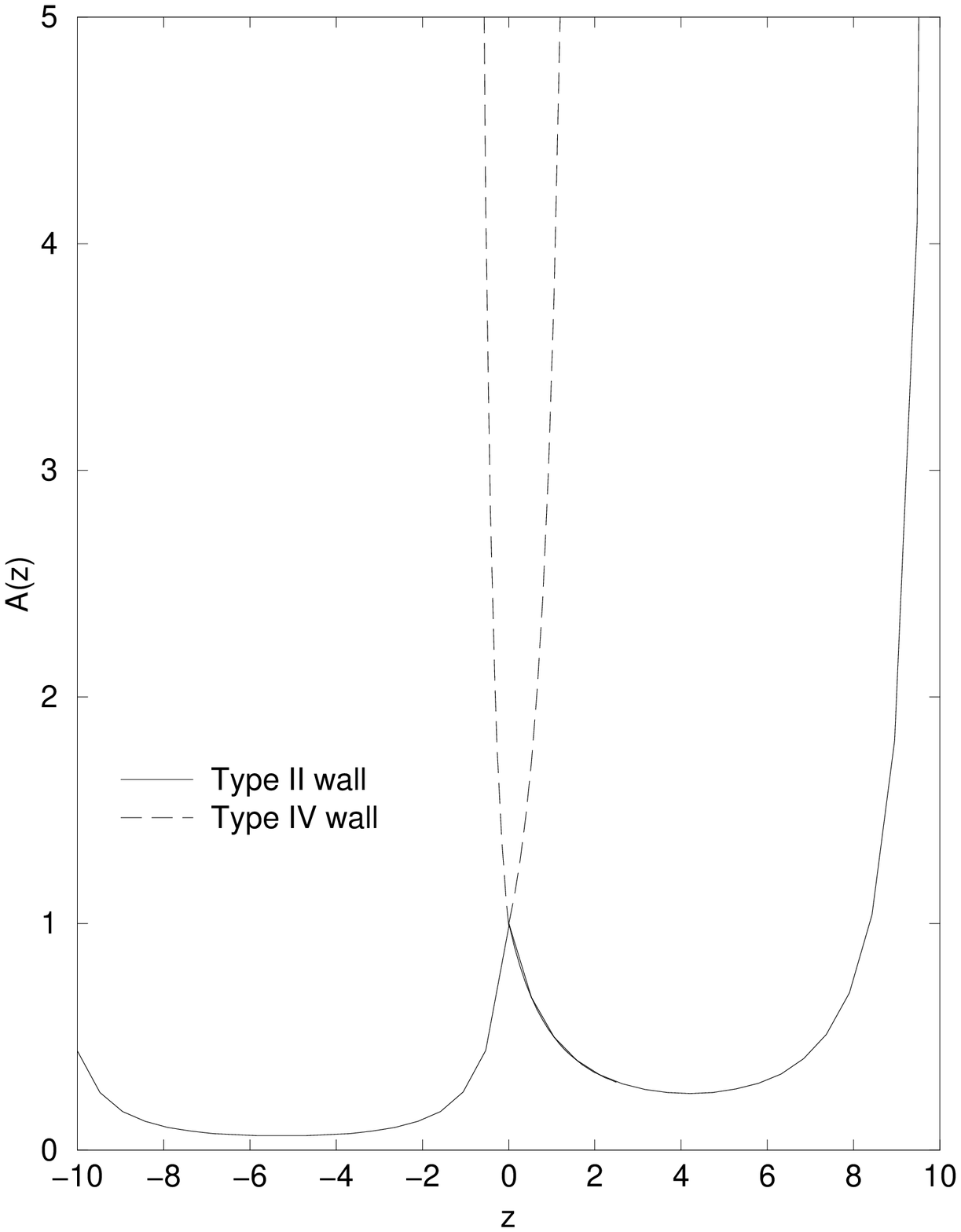}
\hskip 0.25truein
\epsfxsize=1.7truein
\epsfbox[70 32 545 740]{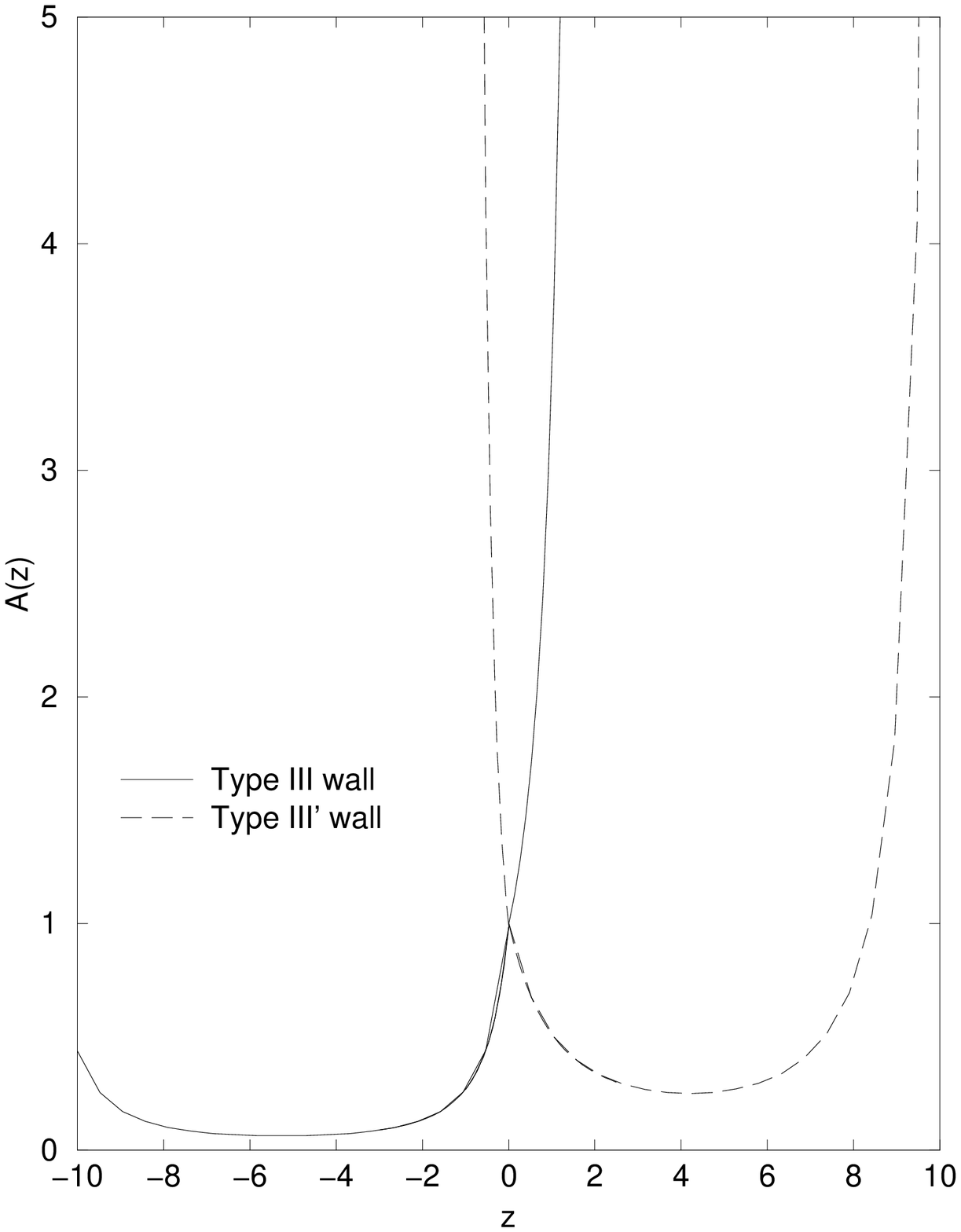}
}
}
\caption{
The metric function $A(z)$ of Type II-IV and Type III-III' walls in the case of ($q_0<0$). They are both AdS-AdS walls with parameter $\alpha_1 = 1$, $\alpha_2 = 1/2$ and $\beta=1/4$. 
}
\end{figure}

\section{Discussion}
\label{discussion}

We have provided a systematic analysis of the  space-time structure
in the background of  infinitely thin vacuum domain walls 
[(D-2)-configurations]  in D-dimensional general relativity. 
We have  shown that the
 homogeneity and isotropy of the space-time intrinsic  to the wall
strongly
 constrains the nature of the space-time (both intrinsic and transverse to
 the well) and that this space-time structure is universal 
for all D-dimensions. The analysis also  revealed  an inherent 
connection between the global and local space-time structure 
of  the wall  and the value of the wall tension
 relative to the  cosmological constants on either side of the wall. 
 
The solutions fall into three classes according to the value of  the 
``non-extremality parameter'' $q_0$: ${q_0}=0$, ${q_0}>0$,  and 
 $q_0<0$. Within each class,
 depending on  whether  gravity is repulsive or attractive near either side
of the wall,  the walls can have positive tension solutions 
(Type I, II, III walls) and negative tension solutions 
(Type III$'$, IV, V walls) whose space-times transverse to the wall display 
  complementary  properties. In this sense
Type I-V, II-IV and III-III$'$  walls  can be viewed as
dual.  (In particular, Type II walls provide a set-up for Randall-Sundrum
scenario  in D=5 with  repulsive gravity  on either side of the wall.)

$q_0=0$ solutions are planar, static configurations. The precise
tuning of their
energy-density to cancel the value of cosmological constant is  ensured by 
supersymmetry.  Such walls exist only for non-positive cosmological
constants. 

Solutions with  positive non-extremality parameter (${q_0}=\beta^2>0$)  are
 expanding ``bubbles'' with 
the space-time internal to the wall  corresponding to the 
expanding de~Sitter (dS$_{D-1}$)  FLRW universe. In particular, 
  Type II walls are 
expanding bubbles with {\it two insides} and  thus ``safe walls'',  Type
III and III$'$ walls 
are bubbles with {\it one inside and one outside} and which sweep out one side of the wall through ``false-vacuum'' decay, 
while Type IV walls  are expanding
bubbles with  {\it two outsides} and thus sweep out the vacuum on either side 
of
the wall. These solutions exist both for positive and negative values of the
cosmological constants.

Solutions with $q_0 =-\beta^2<0$
describe  an anti-deSitter (AdS$_{D-1}$) FRWL universe internal to 
the wall. However, the  coordinates describe 
only a patch of the AdS space-time with the coordinate singularities
 at $t=\pm {\pi\over {2\beta}}$. These walls have solutions
only for the  negative values of  cosmological constants, and do not have 
cosmological horizons in directions transverse to the wall. (Their
energy density is complementary to that of  walls with $q_0>0$.) 
 Further investigations of the geodesics
extensions and their global structure  is needed.

While the work provides a classification of  vacuum domain
wall space-times, we did not address in detail the dynamic issues  such as
their stability  or the  nature of their creation, nor did we 
elaborate on a field-theoretic embedding of such domain walls.
Let us mention again  that AdS-AdS Type II
 walls with $q_0>0$ may  be realized via quantum cosmology 
\cite{QuantumCosmology}
 and that Type III walls are Euclidean  bounce solutions of false-vacuum
 decay bubbles.
As for field-theoretic realization,  positive tension extreme 
walls could be realized as bosonic 
configurations in supersymmetric theories, corresponding to a  kink
solution 
interpolating between two supersymmetric minima.  However, it is expected 
that negative tension walls
 are unstable due to the appearance of 
 a ghost mode.~\footnote{We would
 like to  thank R. Sundrum for a discussion on this point.} 
 The gauged supergravity solutions tend to provide a frame-work for 
negative
 tension extreme wall solutions, i.e. the kink solution interpolates
between 
 {\it supersymmetric maxima}.  This
 issue requires further study and it may have a resolution
 in the string theory context (see also, e.g, 
 \cite{deBVV}).

 The domain wall solutions studied in this paper can  be 
 stacked-up   in the transverse $z$-direction,  
thus provide a solution for an array of parallel walls.   
 In particular, if  D-dimensional space-time has
 possible vacuum solutions with  cosmological constants 
 $\Lambda_{1,2,\cdots n}$,  then one can superimpose  in $z$ direction
  different types of domain walls interpolating between these vacua; this
 may  yield interesting 
possibilities with   phenomenological implications. 
However, the field-theoretic embedding of  such multi-wall set-ups
 may be difficult; the multi-kink solutions are 
supposed to interpolate continuously between (isolated) extrema of the
potential and the desired solution may not exist.

Let us consider specific examples with static (extreme) walls in D=5.
Extreme Type II walls provide a set-up for Randall-Sundrum
scenario with one positive tension brane in D=5 with repulsive gravity 
 on either side of the wall \cite{070}.  
The scenario with one positive tension brane and one negative tension 
brane   \cite{050}, can be realized as a special ($Z_2$-symmetric)
  periodic array of Type II and Type IV extreme wall. 
On the other hand, the realization of such an array within field
theory  may
    be hard to realize  and  it should clearly involve more than one
scalar.
The example of \cite{240} is  a superposition of Type II
 and Type III wall.  It could be realized with a scalar field  that
 interpolates between two supersymmetric AdS 
minima with large enough potential
 barrier which yields Type II wall ($\kappa_D\sigma=\alpha_1+\alpha_2$),
and
the third deeper minimum with a potential barrier insufficiently
 large  yields Type III wall ($\kappa_D\sigma =\alpha_3-\alpha_2$).
   (Note  however, that in spite of its positive energy-density the
   Type III walls are inherently unstable; for nonzero extremality
parameter
   $q_0>0$ they turn into false vacuum decay
    bubbles, sweeping out the space  on one side of this wall.) 
 Another interesting possibility is a superposition of the two  extreme 
 Type I walls \cite{Wangetal}, which can  be realized  via a single
 kink and  anti-kink   that interpolate between anti-deSitter 
     and  Minkowski  supersymmetric minima.

The non-static Type II walls in AdS$_5$ both in the case of $q_0>0$ and 
$q_0<0$ are those studied in \cite{170,Kraus,Sasaki,Garriga}. 
Another intriguing possibility may be a superposition
 of these solutions with  $q_0>0$ (or $q_0<0$), where the conformal factors
 can again be matched from one wall to another. Note however, the
 non-static nature of these solutions may involve pathologies of
 space-times such as bubbles of false vacuum decays, and  require further
    investigations. 
    
 We would like to conclude with a few  remarks  regarding the 
  nature of the  non-extremality parameter $q_0=\beta^2>0$ 
    within cosmological context.  (For related ideas
 implemented in the context of AdS/CFT correspondence, see
\cite{Gubser}.)
     Extreme domain walls ($q_0=0$) are static due to the
     ``miracle of supersymmetry''.  Thus,  in a cosmological context,
     at zero temperature ($T=0$), domain walls between supersymmetric
     vacua remain static. On the other hand, at finite temperature $T>0$,
     supersymmetry is broken,  and  thus the domain
     walls  are those with non-zero $q_0=\beta^2$.
Namely temperature corrections to the scalar corrections are
$\propto T^2$, thus modifying the energy density of the wall
$\sigma=\sigma_{ext}+{\cal O}(T^2)$.  Clearly, since the leading
corrections to $\sigma$ are of ${\cal O}(q_0=\beta^2)$ the result
implies that $q_0\propto T^2$ (or $\beta\propto T$).
     In particular,  the  static extreme Type II  [Type III] domain wall 
(at $T=0$)
      becomes  a non-extreme Type II [Type III] solution (at $T>0$)  
      which is the expanding de-Sitter FLRW bubble with
      two  [one] insides. Thus the  positive cosmological constant 
      intrinsic to the wall as well
      as the rate of expansion of the bubble  are 
      proportional to $\beta\propto T$.
      Thus as the universe cools the  expansion rate and the 
      cosmological constant on the wall  decrease.

\section*{Acknowledgments}

This work was supported in part by
U.S.\ Department of Energy Grant No.~DOE-EY-76-02-3071 (M.C.), 
No.~DE-AC02-76CH03000 (J.W.) and   in part by the University  
of Pennsylvania Research Foundation award (M.C.). We would like to thank K.
Behrndt, J. Lykken and R. Sundrum for discussions.





\end{document}